\documentclass[11pt]{article}
\pdfoutput=1
\usepackage{jheppub}
\usepackage{multirow,soul}

\title{\boldmath A cosmological case study of a tower of warm dark matter states: $N$naturalness}
\author[\$,a]{Saurabh Bansal,}
\author[b,c]{Subhajit Ghosh,}
\author[d]{Matthew Low,}
\author[c]{and Yuhsin Tsai}
\affiliation[\$]{A large financial institution}
\affiliation[a]{Department of Physics, University of Cincinnati, Cincinnati, Ohio 45221, USA}
\affiliation[b]{Texas Center for Cosmology and Astroparticle Physics, Weinberg Institute,
Department of Physics, The Unversity of Texas at Austin, Austin, TX 78712, USA}
\affiliation[c]{Department of Physics, University of Notre Dame, IN 46556, USA}
\affiliation[d]{PITT PACC, Department of Physics and Astronomy,\\ University of Pittsburgh, 3941 O’Hara St., Pittsburgh, PA 15260, USA}

\emailAdd{bansalsh@ucmail.uc.edu}
\emailAdd{sghosh@utexas.edu}
\emailAdd{mal431@pitt.edu}
\emailAdd{ytsai3@nd.edu}
\preprint{UTWI-31-2024}

\abstract{In this work, we study the cosmological effects of a tower of warm dark matter states on the cosmic microwave background (CMB) and on large-scale structure (LSS). For concreteness, we consider the $N$naturalness model, which is a proposed mechanism to solve the Higgs hierarchy problem.  In this framework, the sector of particles of the Standard Model is copied $N$ times where the Higgs mass-squared value is the only parameter that changes between sectors.  The other sectors are similar to our own, except their particles are proportionally heavier and cooler compared to the Standard Model sector.  Since each sector is extremely weakly coupled to other sectors, direct observations of the new particles are not expected.  The addition of new photon-like species and new neutrino-like species, however, can be detected through the CMB and in LSS data. These additional neutrinos form a tower of states with increasing mass and decreasing temperature compared to the SM neutrinos. This tower causes a more gradual suppression of the matter power spectrum across different comoving wavenumbers than a single warm dark matter state would.  We quantitatively explore these effects in the $N$naturalness model and compute the parameter space allowed by the Planck 2018, weak lensing, and Lyman-$\alpha$ datasets. Depending on the underlying parameters, Planck 2018 and weak lensing data can require $N$naturalness to be tuned at the $10\%$ level for Dirac neutrinos and at the $5\%$ level for Majorana neutrinos.  The additional neutrino states are crucial in constraining the model, particularly through their suppression of the power spectrum at small scales.  The inclusion of many warm dark matter species is computationally very challenging and we make detailed assessments of our approximations and comment on potential future improvements.}

\begin{document}

\maketitle
\flushbottom

\section{Introduction}
\label{sec:introduction}

The early Universe is a unique laboratory for exploring new physics across a broad range of energy scales.  At extremely high energies, new physics occurring during inflation can modify curvature perturbations and imprint signatures on the cosmic microwave background (CMB) and large-scale structure (LSS).  On the other hand, at relatively low energies, new physics near the MeV scale can be constrained through measurements of the number of additional relativistic degrees of freedom during Big Bang Nucleosynthesis.  Even lower, below the keV scale, new self-interactions among dark sector particles can leave imprints on the CMB and structure formation.  Finally, below 100 eV, the temperature spectrum of the CMB is sensitive to changes from the expected behavior of the photon-baryon plasma in the form of energy injections or leakage.  Notably, these scales bookend rather than correspond to the scale of the most interest for particle physics, namely the electroweak scale of 100 GeV.

The electroweak scale harbors the enduring mystery of the Higgs hierarchy problem which is the question of why the scale of the Higgs is at 100 GeV when, in the absence of new physics, its more natural value is at the Planck scale.  This problem has, in large part, motivated many of the searches at the Large Hadron Collider (LHC), and the subsequent model building to account for the absence of new particles at LHC-accessible energies~\cite{Batell:2022pzc}.  Even in models where there are no new particles at the TeV scale, the dynamics that allow for a light Higgs, still dominantly occur at the TeV scale.  For this reason, the conventional knowledge is that cosmology is not a sensitive probe of models that address electroweak naturalness.

Counter to this expectation, however, there are several models that both solve the Higgs hierarchy problem and are testable, in a meaningful way, using cosmology.  The common feature of such models is a correlation between the TeV scale dynamics that address naturalness and new particles at lower energies that can leave imprints on the CMB or in LSS.  In this regard, cosmology not only probes these scenarios but even makes quantitative statements about how tuned the remaining parameter space is.

One example of such a model is the Mirror Twin Higgs model~\cite{Chacko:2005pe}, where the cosmological features were described in Refs.~\cite{Craig:2016lyx,Chacko:2018vss,Chacko:2021vin} and detailed numerical predictions were made in Refs.~\cite{Bansal:2021dfh,Bansal:2022qbi,Zu:2023rmc}.  This model concretely predicts a spectrum of mirror sector particles with all cosmological properties fixed by the temperature of the mirror sector, the baryon abundance of the mirror sector, and the tuning of the model.\footnote{The mirror temperature can differ from that of the Standard Model sector due to an asymmetric reheating process~\cite{Craig:2016lyx,Chacko:2016hvu,Ireland:2022quc}, which can be sourced by the production of asymmetric twin baryons~\cite{Koren:2019iuv,Bittar:2023kdl,Alonso-Alvarez:2023bat}.}  Several cosmological signatures are predicted, including dark acoustic oscillations between mirror baryons and mirror photons and the presence of both free-streaming and interacting radiation in the form of mirror neutrinos and mirror photons~\cite{Chacko:2018vss,Bansal:2021dfh}.  The mirror symmetry of the model ties the mirror particle dynamics near 10 eV to the Higgs hierarchy problem near 100 GeV which allows CMB and LSS data to set constraints on such models.  While the Mirror Twin Higgs model is one case study, the results and lessons apply far more generally because the low-energy ingredients of the model are very universal.  New particles with properties similar to those of the Standard Model are added, but with different masses, different abundances, and different temperatures.

In this work, we explore the cosmology of another solution to the Higgs hierarchy problem which is the $N$naturalness model~\cite{Arkani-Hamed:2016rle}.  In a nutshell, $N$naturalness solves the hierarchy problem by the inclusion of a large number of sectors, copies of the Standard Model, with different values of the Higgs mass-squared parameter.  These sectors both lower the effective gravity scale and lower the naively expected Higgs mass which solves the hierarchy problem.  The primary cosmological consequence of this setup is a tower of Standard Model-like particles which become colder as they become heavier.  The most cosmologically relevant new particles are the additional photons (the ``$N$-photons''), which act as dark radiation, and the additional neutrinos (the ``$N$-neutrinos''), which act as warm dark matter (DM).

The two main parameters that determine the cosmological properties of the $N$naturalness model are: firstly, the mass of a particle, called the reheaton, that is responsible for reheating the Universe and distributing its energy density among the different sectors and secondly, the mass scale of the new sectors which is also a proxy for the tuning of the model.  The distribution of energy density among different sectors leads to a temperature profile among the sectors such that the heavier a sector is, the colder it is.\footnote{This is only true for the ``Standard Model like'' sectors.  The ``exotic'' sectors do not follow this pattern, but their energy density is sufficiently low as to be negligible over the majority of parameter space.}  With the mass scale and temperature of each sector set by the model parameter choices, the cosmological consequences can be calculated.

In early works, the primary observable considered was the value of $\Delta N_{\rm eff}$ from the $N$-photons~\cite{Arkani-Hamed:2016rle,Choi:2018gho}.  It was also predicted that the tower of massive neutrinos would suppress the matter power spectrum~\cite{Banerjee:2016suz}, but the signature was not studied in detail.  Some of the sectors could have a first-order QCD phase transition which was recently shown to be potentially observable in gravitational wave observatories~\cite{Batell:2023wdb} (see also Ref.~\cite{Archer-Smith:2019gzq}).  Other ideas have also been explored~\cite{Baumgart:2021ptt,Easa:2022vcw}.  In this work, we compute the full cosmological history of $N$naturalness and compute parameter posterior distributions from data.  As we will describe, the thermal history of the $N$-neutrinos, in particular, can be quite complex and will be seen to have a significant impact on the cosmological bounds.  The addition of many warm DM species substantially alters the CMB and LSS signals, which allows us to set much stronger constraints on $N$naturalness than $\Delta N_{\rm eff}$ alone. 

While we arrive at our cosmological model of multiple warm DM states through seeking a solution for the hierarchy problem such a setup is possible in simplified scenarios.  In fact, consider the SM itself, with a normal mass hierarchy, one massless neutrino and two massive neutrinos already make a contribution both to radiation and matter.  Studies have shown that the difference in the distribution of neutrino masses between the normal and inverted hierarchies has observable effects on the matter power spectrum~\cite{Agarwal_2010}.  With only the SM, however, the upper mass limit from cosmological observations of the neutrinos is well established which means the SM neutrinos are necessarily still relativistic at the smallest scales that the CMB can probe, which limits the possible effects on the matter power spectrum.  The addition of new sterile neutrinos, however, allows for species that transition from being relativistic to non-relativistic over scales that are probed by the CMB, leading to a wider range of modifications to the matter power spectrum.  This simple scenario aligns more closely with the usual warm DM parameter space.  In models of warm DM, many extensions of a single new species have been studied, such as non-thermal distributions~\cite{Cuoco:2005qr,Acero:2008rh} which could be induced through decays between multiple species.  The possible effects on the matter power spectrum are rich in such scenarios~\cite{Dienes:2020bmn,Dienes:2021cxp,Dienes:2021itb}.

In general, the cosmological signatures of models with multiple warm DM species -- as opposed to single-species warm DM scenarios -- are fairly under-explored.  The presence of a tower of warm DM states in particular, induces novel features, such as scale-dependent $\Delta N_{\rm eff}$ and scale-dependent suppressions of the matter power spectrum.  These have not been systematically studied in the literature.  In this work, we use $N$naturalness as a concrete example that contains a tower of neutrino states.  As we will show, compared to a model of a single warm DM species that produces the same $\Delta N_{\rm eff}$ and dark matter abundance today, a tower of warm DM states results in a suppression of the matter power spectrum at comparatively smaller length scales.  This leads to a distant shape of the matter power spectrum that cannot be replicated with a single warm DM species.

While our study focuses on $N$naturalness, which is one specific model of multiple warm DM species, that has a particular relation between the mass and temperature of additional neutrino species, the cosmological effects are quite similar to other models as long as they have multiple species of warm DM that become non-relativistic at different times.  Even though in principle, freely varying the temperature and mass of additional neutrinos species allows for a wide array of cosmological modifications, in practice the tight bound on $\Delta N_{\rm eff}$ and the precise measurement of the dark matter abundance only admits a subset of these models and such models tend to resemble $N$naturalness in terms of their temperature and mass profiles.

One challenge that is common between models of multiple warm DM species is the significant computational power required to solve the phase space integral for each species of warm DM.  Typically as a matter of tractability, it is necessary to truncate the neutrino tower and only include the lightest, and most cosmologically relevant, states.  The higher the comoving wavenumber $k$ of a perturbation that we want to calculate, the more warm DM species that must be included to reliably predict the signal.  In our work, we primarily focus on observables with $k \lesssim 0.3~{\rm Mpc}^{-1}$ and truncate the neutrino tower appropriately based on the masses and kinetic energy evolution.  One dataset that we make an exception for is Lyman-$\alpha$ data where we extend our calculation to $k = 1~{\rm Mpc}^{-1}$ and include additional sectors as is needed.

The paper is organized as follows.  In Sec.~\ref{sec:model} we review the $N$naturalness model and its parameters as well as the behavior of the $N$-photons.  We describe the implementation of the model in \texttt{CLASS} in Sec.~\ref{sec.class}, and discuss the behavior of CMB and matter power spectrum due to the presence of the $N$-neutrinos in Sec.~\ref{sec:neutrinos}. In Sec.~\ref{sec:N-nu}, we study the difference in matter power spectrum suppression between one and $N$ warm DM species.  We conduct the Markov-Chain Monte Carlo (MCMC) analysis in Sec.~\ref{sec:mcmc} and show that the inclusion of $N$-neutrino dynamics strengthens the $N$naturalness constraint significantly. In Sec.~\ref{sec:lyalpha}, we estimate the sensitivity of Lyman-$\alpha$ constraints on the model. We conclude in Sec.~\ref{sec:conclusion}.  Appendix~\ref{app:tables} contains additional plots and tables.

\section{The Model}
\label{sec:model}

In this section, we describe the general features of models of $N$naturalness and the specific model we use in our study.

\subsection{Sectors}

In the $N$naturalness framework there are $N$ sectors of particles.  Each sector, labeled by $i$, is assumed to be the same except for its $m_{H,i}^2$ parameter.  If these parameters are allowed to vary with uniform spacing, their values are
\begin{equation} \label{eq:mh2}
m_{H,i}^2 = - \frac{\Lambda_H^2}{N}(2i+r),
\quad\quad\quad
-\frac{N}{2} \leq i \leq \frac{N}{2},
\end{equation}
where $\Lambda_H$ is the cut-off.  The Standard Model is identified to be the $i=0$ sector with $m_H^2 = m_{H,0}^2 = - (88~{\rm GeV})^2$.  The parameter $r$ controls how close the sector with the smallest negative $m_H^2$ value is to zero and is a good proxy for the amount of fine-tuning in the model.  When $r \ll 1$ the theory is considered tuned, meaning that it does not address the hierarchy problem well, because the cut-off is $1/\sqrt{r}$ larger than the naive expectation.  Equivalently, when $r \ll 1$, $m_{H}^2$ is closer to zero than naively expected.

The inclusion of $N$ sectors has two immediate consequences.  Firstly, it renormalizes the scale where gravity becomes strong to $M_{\rm Pl}/\sqrt{N}$.  Secondly, due to Eq.~\eqref{eq:mh2}, the lightest negative $m_H^2$ parameter is approximately $\Lambda_H^2/N$, which is substantially lower than the squared cut-off $\Lambda_H^2$ when $N$ is large.

Choosing $N \sim 10^{16}$ would raise $\Lambda_H$ to $\sim 10^{11}~{\rm GeV}$ while simultaneously lowering the gravitational scale to the same value, thus solving the full hierarchy problem.  Smaller values of $N$ would leave a gap between $\Lambda_H$ and the gravitational scale and would require another fine-tuning solution, like supersymmetry, between those scales.  The value of $N \sim 10^4$ is a natural benchmark because it raises $\Lambda_H$ enough to solve the little hierarchy problem while pushing the gravitational scale down to coincide with the scale of grand unification~\cite{Arkani-Hamed:2016rle}.

Sectors with $m_H^2 < 0$ experience electroweak symmetry breaking in the same way as the SM and we call these SM-like sectors.  The more negative the value of $m_{H,i}^2$ in an SM-like sector $i$, the larger the vacuum expectation value (VEV) $v_i$ of that sector is compared to that of the SM $v_{\rm SM}$.  The particle spectrum, with the possible exception of neutrinos, in an SM-like sector $i$ is the same as the SM but with masses that are larger by a factor of $v_i / v_{\rm SM}$.  Dirac neutrinos follow this scaling, but Majorana neutrinos are an exception and scale as $(v_i / v_{\rm SM})^2$.

Sectors with $m_H^2 > 0$, on the other hand, do not experience electroweak symmetry breaking via the Higgs potential.  Instead, in these sectors, the chiral condensate of the corresponding QCD sector breaks electroweak symmetry at a much lower scale of $f_\pi \sim 100~{\rm MeV}$.\footnote{Due to the differences in quark masses, the running of $\alpha_s$ in each sector is slightly different resulting in slightly different values of $f_\pi$.  The variation, however, is very mild so we refer to a single value of $f_\pi$ for simplicity.}  The massive vector bosons have masses of $\sim g f_\pi \sim 100~{\rm MeV}$, rather than $\sim 100~{\rm GeV}$.  A Higgs VEV is induced through the couplings of the Higgs to the chiral condensate which leads to very light fermion masses of $\sim y_t y_f \Lambda_{\rm QCD}^3 / m_{H,i}^2$, where $y_f$ is the Yukawa coupling of a fermion $f$.  We call these exotic sectors since their dynamics and cosmology are different from that of the SM.

\subsection{Reheating}

In order to have a viable cosmological history, the SM sector must be dominantly populated.  This is accomplished through a particle called the reheaton which is assumed to dominate the energy density of the universe at some early time.  The reheaton then decays into all available channels, dominantly populating the SM, but also leaving potentially measurable amounts of energy density in the other sectors.

In this work, we consider the scalar reheaton model where a new real scalar $\phi$ is added
\begin{equation}
\mathcal{L}_\phi \supset 
\frac{1}{2} \partial_\mu \phi \partial^\mu \phi
- \frac{m_\phi^2}{2} \phi^2
- a_\phi \phi \sum_i |H_i|^2.
\end{equation}
The mass of the reheaton is $m_\phi$ and $a_\phi$ is a trilinear term that couples $\phi$ and two Higgs particles from each sector with the same magnitude for all sectors.\footnote{To maintain a well-behaved theory in the large $N$ limit $a_\phi$ should have an arbitrary sign for each sector with a magnitude that scales as $|a_\phi| \lesssim \Lambda_H / N$.} 

The trilinear term results in mixing between each Higgs that undergoes electroweak symmetry breaking and $\phi$.  The mixing between the Higgs of the SM sector and $\phi$ leads to an upper bound on $a_\phi$ from Higgs mixing constraints of roughly $|a_\phi| \lesssim 1~{\rm MeV}$~\cite{Arkani-Hamed:2016rle}.  Only ratios of couplings enter into cosmological observables making these observations nearly insensitive to the value of $a_\phi$.

Decays into SM-like sectors therefore proceed via mixing with the Higgs of that sector.  Partial widths scale as $\Gamma_i \sim 1 / m_{h,i}^2$ and the number of kinematically open decay channels decreases as $i$ increases because the mass scale of SM-like sectors gets higher and higher.  The energy density $\rho_i$ in a sector $i$ will scale as $\rho_i / \rho_{\rm SM} \approx \Gamma_i / \Gamma_{\rm SM}$ which means that only a handful of sectors have a relative energy density larger than the per mille level.  The total energy density, however, is approximately logarithmically sensitive to $i$ until $i$ is large enough that the number of open decay channels drops substantially.

At the scale of the reheaton mass, electroweak symmetry in the exotic sectors is unbroken so there is no mixing between the reheaton and each Higgs of the exotic sectors.  Decays into these sectors, instead, proceed dominantly through a loop of Higgses into a pair of vectors~\cite{Arkani-Hamed:2016rle}.  When $m_H < m_\phi < 2m_H$ the three-body decay of $\phi \to H t_R t_L$ is possible but is only numerically relevant when $m_\phi$ is very close to $2m_H$~\cite{Batell:2023wdb}.  The loop decay leads to a partial width that scales as $\Gamma_i \sim 1 / m_{H,i}^4$.  The energy density in the exotic sectors is negligible in most of the parameter space.  If the range of $r$ is extended to larger than $1$ then there are small regions where the $i=-1$ can have a measurable energy density~\cite{Batell:2023wdb}.

\subsection{Parameter Space}

The theory is characterized by the number of sectors $N$, the reheaton mass $m_\phi$, the spacing parameter $r$, the trilinear coupling $a_\phi$, and whether the neutrinos are Majorana or Dirac.  The properties of $N$naturalness cosmology, however, are determined by a subset of these parameters: the continuous parameters $m_\phi$ and $r$, the binary choice of Majorana or Dirac neutrinos, and a discrete set of values of $N_\nu \leq N/2$. 

Since cosmological observables depend on the ratio of reheaton partial widths they are almost entirely insensitive to the value of $a_\phi$.  For concreteness we set $a_\phi = 1~{\rm MeV}$.  The ratio of reheaton partial widths does depend on both $m_\phi$ and $r$.  We consider values of $m_\phi$ between 5 GeV and 300 GeV.  Above 300 GeV the reheaton can start to decay on-shell to pairs of SM-like Higgses in the lightest sectors which disrupts the $N$naturalness mechanism.  For $r$, we consider values between 0 and 1.\footnote{In Ref.~\cite{Batell:2023wdb} it was shown that the range of $r$ can be consistently extended up to 2.} 

Of the new particles added in $N$naturalness, the cosmologically relevant ones are the photons from the additional sectors and the neutrinos from the additional sectors.  We refer to the collection of these from all new sectors, respectively, as the $N$-photons and $N$-neutrinos.  For the $N$-photons we fix the number of sectors to $N=10^4$  which includes $N/2$ SM-like sectors and $N/2$ exotic sectors.  This value addresses the little hierarchy problem and is consistent with overclosure bounds from the total density of cold dark matter (CDM)~\cite{Arkani-Hamed:2016rle}.  Results are very insensitive to the value of $N$ as sectors with large $|i|$ have very suppressed contributions to $\Delta N_{\rm eff}$.

Regarding the $N$-neutrinos, we only include the neutrinos from $N_\nu$ SM-like sectors, where $N_\nu \leq N/2$.  Computationally it is intractable to solve the Boltzmann equations for such a large number of neutrino species so we aim to use the smallest value of $N_\nu$ that leads to accurate results.  We are able to perform calculations for values of $N_\nu \lesssim 50$, but we study in detail the estimated impact of truncating the heavy neutrino tower. To our advantage, we will show that neutrinos from sectors with an index $i\gtrsim 100$ can be safely treated as CDM for even the smallest scale cosmological measurements.  In most cases, we use $N_\nu = 20$ as the default benchmark, with the exception of our Lyman-$\alpha$ analysis which requires $N_\nu = 50$.

Finally, in all cases, we consider Majorana and Dirac neutrinos separately.  Their behavior differs because the mass scale of the Majorana neutrinos grows much faster.  As a consequence, Majorana neutrinos create larger suppressions in the matter power spectrum at small scales and Majorana neutrinos contribute more energy density at late times.

\subsection{$N$-photon Signals}

The impact of the $N$-photons on cosmological observables is primarily encapsulated by the contribution to the number of relativistic degrees of freedom, $\Delta N_{\rm eff}$.  The photons from the SM-like sectors are free-streaming and their contribution can be approximated as 
\begin{equation} \label{eq:dneff_i>0}
\Delta N_{{\rm eff},i>0} = 
\frac{8}{7} \left(\frac{11}{4}\right)^{4/3}
\frac{1}{\Gamma_{\rm SM}}\sum_{i>0} \Gamma_i,
\end{equation}
where $\Gamma_{\rm SM}$ is the partial width of the reheaton decaying into all kinematically accessible states of the SM and $\Gamma_i$ is the partial width of the reheaton decaying into all kinematically accessible states of sector $i$.  Eq.~\eqref{eq:dneff_i>0} neglects the mild dependence that arises from changes in the degrees of freedom $g_{*,i}$ in a sector $i$ as a function of temperature.  This was calculated in Ref.~\cite{Batell:2023wdb} and found to only introduce $\mathcal{O}(20\%)$ corrections to the $\Delta N_{\rm eff}$ calculation over the model parameter space we study.

The photons from the exotic sectors may be interacting at the time of the CMB because the corresponding electrons are much lighter and stay in thermal equilibrium with their respective photons until much later times.  Additionally, the exotic sectors likely undergo first-order QCD phase transitions\footnote{Earlier studies, {\it e.g.} Refs.~\cite{PhysRevD.29.338,Butti:2003nu,Iwasaki:1995ij,Karsch:2003jg}, have suggested that QCD with vanishing quark masses would undergo a first-order phase transition, but some modern lattice results disfavor this claim~\cite{Cuteri:2021ikv}.} which result in an increase in energy density from the phase transition~\cite{Batell:2023wdb}.  For a phase transition of strength $\alpha$, which is defined as the ratio between the change of QCD vacuum energy change and the total energy density in each sector, the energy density increases roughly by a factor of $1+\alpha$, where typically values of $\alpha$ may range from $0.05 - 10$~\cite{Helmboldt:2019pan,Bigazzi:2020avc,Reichert:2021cvs}.  In this work, we safely neglect these effects as our parameter space does not contain regions where the energy density from exotic sectors is non-negligible.\footnote{To validate our statement that the exotic sector energy density is negligible we can compare $\Delta N_{{\rm eff},i>0}$ from Eq.~\eqref{eq:dneff_i>0} with the same summation over the exotic sectors $\Delta N_{{\rm eff},i<0}$.  When the ratio of $(\Delta N_{{\rm eff},i<0})/(\Delta N_{{\rm eff},i>0})>1$ the value of $\Delta N_{\rm eff} > 0.58$.  For ratios of $0.1, 0.01, 0.001$ the corresponding limits on $\Delta N_{\rm eff}$ are $0.22, 0.13, 0.09$, respectively.  Therefore any parameter space where the relative energy density in the exotic sectors is large also has a large value of $\Delta N_{\rm eff}$.}

\begin{figure}
  \centering
  \includegraphics[width=0.85\textwidth]{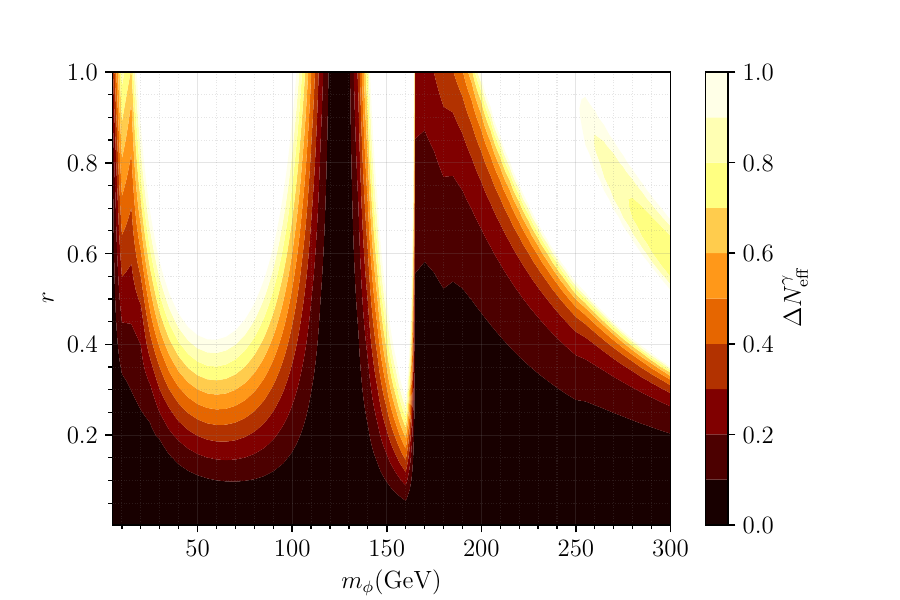}
  \caption{The number of effective relativistic degrees of freedom, only from photons, $\Delta N_{\rm eff}^\gamma$, in the plane of $r$ and $m_\phi$.}
  \label{fig:dneff_gamma}
\end{figure}

In our study, we use $\Delta N_{\rm eff}^\gamma = \Delta N_{{\rm eff},i>0}$.  The values of $\Delta N_{\rm eff}^\gamma$ in the $(m_\phi, r)$ parameter space are shown in Fig.~\ref{fig:dneff_gamma}.
We include the three-body and four-body decays of the reheaton to two vectors and to two scalars.
This is the only bound directly considered in Ref.~\cite{Arkani-Hamed:2016rle} and serves as a useful benchmark to compare the impact that constraints from other cosmological observables, especially neutrinos, have on the $N$naturalness parameter space.  The current bound from the CMB is $\Delta N_{\rm eff}\lesssim 0.5$ which includes data from the SH0ES collaboration~\cite{Planck:2018vyg,Blinov:2020hmc,Riess:2021jrx}.  
The naive $\Delta N_{\rm eff}^\gamma$ constraint, for most $m_\phi$ values, requires $r\lesssim 0.2-0.4$, corresponding to a mild $20-40\%$ tuning in the model. 

The bounds become significantly weaker in three $m_\phi$ regions either due to a larger value of $\Gamma_{\rm SM}$ or a smaller value of $\sum_i \Gamma_i$, as seen in Eq.~\eqref{eq:dneff_i>0}:.  These regions are the following.  (1) Small $m_\phi \lesssim 20$ GeV. Here, $m_\phi$ decays predominantly into fermions.  Decays of the reheaton to two SM bottom quarks are open, but depending on $r$, many of the decays to $i>0$ bottom quarks become kinematically inaccessible.  (2) $m_\phi$ close to 125 GeV which is the mass of the SM Higgs. In this case, the reheaton has substantial mixing with the SM Higgs, resulting in dominant decay into the SM sector. (3) $m_\phi$ above $\approx 160$ GeV, which is twice the mass of the SM $W$ boson. Here, the reheaton can decay into two SM $W$ bosons with a large branching ratio. 

\section{Implementation in \texttt{CLASS}}
\label{sec.class}

In order to calculate the detailed cosmological signals of the $N$naturalness model, we have modified the Boltzmann solver code Cosmic Linear Anisotropy Solving System (\texttt{CLASS})~\cite{Diego_Blas_2011}.  In this section, we comment on the subtleties and the approximations of our \texttt{CLASS} implementation.

We have assumed a negligible baryon asymmetry in all SM-like and exotic sectors.  This is a motivated choice both in light of overclosure bounds and in light of baryogenesis considerations~\cite{Arkani-Hamed:2016rle}.  Relaxing the constraint will imply that part of the total CDM today is constituted by the baryons and leptons of the SM-like and exotic sectors. The cosmological effects of the additional sectors are, therefore, manifested through photons and neutrinos in the additional sectors.  Roughly speaking, the additional photons contribute as dark radiation and the additional neutrinos contribute as warm dark matter when relatively light and as CDM when relatively heavy.  These contributions are calculated for each point in the $N$naturalness parameter space.

The additional photons impact cosmological evolution through their contribution to $N_{\rm eff}$ which we denote as $\Delta N_{\rm eff}^\gamma$.  We calculate $\Delta N_{\rm eff}^\gamma$ analytically outside of \texttt{CLASS} which includes a summation over all SM-like sectors.

The impact of an additional species of neutrino depends on the mass of the species.  When it is sufficiently heavy, it acts as CDM and contributes to the CDM abundance.  At lighter masses, however, each neutrino species acts as warm dark matter and its impact should be included by solving the momentum-dependent Boltzmann equations.  Therefore, we should be able to include a subset of the additional neutrino species in the Boltzmann equations and adjust the energy density of CDM without sacrificing any accuracy.  As mentioned in the previous section, $N_\nu$ is the number of sectors whose neutrinos are included in the full Boltzmann equations.  Solving the Boltzmann equations for the additional neutrinos requires the temperature and masses of the neutrinos from each sector.  For a given point in parameter space, the neutrino temperatures and masses are input into \texttt{CLASS} through a precomputed table schematically represented in Table~\ref{tab:interp_table}. 

Solving the Boltzmann equations of momentum-dependent massive neutrinos is highly computationally intensive~\cite{Kamionkowski:2021njk}.  For each sector of neutrinos, there are three neutrino species requiring $3N_\nu$ sets of Boltzmann equations.  For $N_\nu \approx N/2 \gg 1$ running the code becomes intractable.  We make the following approximations.  Firstly, we assume that the three species of neutrinos in the sector are degenerate within that sector.  This is achieved in \texttt{CLASS} by setting the flag \texttt{deg\_ncdm = 3}.  The SM neutrinos are also taken to be degenerate with a sum of masses of $\sum m_{\nu,{\rm SM}} = 0.12~{\rm eV}$, which is the well-established $2\sigma$ upper bound from Planck and BAO datasets~\cite{Planck:2018vyg}.  Although degenerate-mass neutrinos are incompatible with neutrino oscillations, the primary cosmological signatures of neutrinos are the sum of neutrino masses and their contribution to $\Delta N_{\rm eff}$.  In particular, the case of three neutrinos per sector $i$ with masses $\{m^{(1)}_{\nu,i} , m^{(2)}_{\nu,i} , m^{(3)}_{\nu,i} \}$ and the case of a single triply-degenerate neutrino per sector $i$ with a mass of $m_{\nu,i} = (m^{(1)}_{\nu,i} + m^{(2)}_{\nu,i} + m^{(3)}_{\nu,i})/3$ are nearly indistinguishable when the total neutrino mass for a sector is $\gtrsim 0.2$ eV which is nearly already the case for our choice of SM neutrino masses~\cite{Lesgourgues:2006nd}.

Secondly, we set precision parameters in \texttt{CLASS} as \texttt{l\_max\_ncdm = 5 } (the default is \texttt{17}) and \texttt{tol\_ncdm\_synchronous = 0.01} (the default is \texttt{0.001}).  We checked this lower precision setting induces $0.6\%$ changes in the \texttt{CLASS} anisotropy spectrum which is below the precision of the Planck data at the corresponding multipole range.  Finally, we do not use $N_\nu = N/2$ with $N = 10^4$, but use a smaller value.  In practice, we use values $N_\nu = 20$ for CMB and $N_\nu = 50$ for LSS (Lyman-$\alpha$) for an accurate approximation of the neutrino effects.  These approximations are studied in detail in Sec.~\ref{sec:neutrinos}.

We summarize the approach below.  As discussed in Sec.~\ref{sec:model}, a point in the $N$naturalness parameter space is parameterized by the values of $r$ and $m_\phi$.  The value of $r$ determines the spacing between the additional sectors and the value of $m_\phi$ impacts the temperature of each sector via the branching ratios of the reheaton.  For each parameter point, we compute  $\Delta N_{\rm eff}^\gamma$ from the $N$-photons from all SM-like sectors and we compute $T_{\nu,i}$ and $m_{\nu,i}$ for each SM-like sector.  These values are stored in a table which is read into \texttt{CLASS}.  Table~\ref{tab:interp_table} shows the schematic form of this table.  When augmented with this table, our modified version of \texttt{CLASS} acts as a two-parameter extension of $\Lambda$CDM.

\begin{table} [tb]
  \centering
  \begin{tabular}{|c|c|c|c|c|c|c|c|c|c|c|c|} \hline
\multirow{2}{*}{$r$} & \multirow{2}{*}{$m_\phi$} & \multirow{2}{*}{$\Delta N_{\rm eff}^\gamma$} & \multicolumn{4}{c|}{ $i=1$} &  \multicolumn{4}{c|}{ $i=2$} & $\cdots$ \\\cline{4-12}
         &  &  & $T_{\nu,1}$ & $ m_{\nu,1}^{(1)}$ & $ m_{\nu,1}^{(2)}$ & $ m_{\nu,1}^{(3)}$ 
    & $T_{\nu,2}$ & $ m_{\nu,2}^{(1)}$ & $ m_{\nu,2}^{(2)}$ & $ m_{\nu,2}^{(3)}$ & $\cdots$ \\
         \hline
  \end{tabular}
  \caption{Schematic form of the table through which $N$naturalness is implemented in \texttt{CLASS}. The $\Delta N_{\rm eff}^\gamma$ encodes the contribution of the $N$-photon for all SM-like sectors. For each sector, the temperature and masses of the neutrinos are computed following Ref.~\cite{Arkani-Hamed:2016rle}. We assume a normal hierarchy for the neutrino mass, however, in \texttt{CLASS} we employ the degenerate neutrino approximation. The table is computed for a dense grid of $r$ and $m_\phi$, and \texttt{CLASS} uses the interpolation technique to calculate quantities for the in-between grid points. Two different tables are used for the Dirac and Majorana cases.}
  \label{tab:interp_table}
\end{table}

\section{$N$-neutrino Signals}
\label{sec:neutrinos}

In this section, we consider the range of signals from the neutrinos of the additional SM-like sectors.  The signals are distinct for the case of Dirac neutrinos and the case of Majorana neutrinos.  While the temperature $T_{\nu,i}$ of the neutrinos of the $i$th sector is independent of the nature of the neutrinos, the neutrino mass, on the other hand, has a different scaling.
Dirac neutrinos have masses that scale as
\begin{equation} \label{eq:dirac}
m_{\nu,i}^{\rm Dirac} 
= \left(\frac{v_i}{v_{\rm SM}}\right) m_{\nu, {\rm SM}}
= \sqrt{\frac{2i+r}{r}} m_{\nu, {\rm SM}},
\end{equation}
where $v_i$ is the VEV of sector $i$, $v_{\rm SM}$ is the VEV of the SM, and $m_{\nu, {\rm SM}}$ is the mass of the corresponding SM neutrino.
For Majorana neutrinos, the scaling is
\begin{equation} \label{eq:majorana}
m_{\nu,i}^{\rm Majorana} 
= \left(\frac{v_i}{v_{\rm SM}}\right)^2 m_{\nu, {\rm SM}}
= \left(\frac{2i+r}{r}\right) m_{\nu, {\rm SM}}.
\end{equation}
We review the cosmological signatures of neutrinos in terms of the mass and temperature. We will study how these signatures change with sector and $N$naturalness parameters using the scaling relations discussed in Sec.~\ref{sec:model}.  We will also see that the signatures are different for Dirac and Majorana neutrinos.

\subsection{Matter Power Spectrum}
\label{sec:mps}

Roughly speaking a given neutrino species will act as cold dark matter when non-relativistic and as radiation when relativistic.  Interesting collective effects emerge due to the tower of neutrino species that span a range of masses compared to the typical warm dark matter scenario with a single species.  In particular, the matter power spectrum is sensitive to the times at which a neutrino species is relativistic or non-relativistic.

\begin{figure}
  \centering
  \includegraphics[width = 0.47\textwidth]{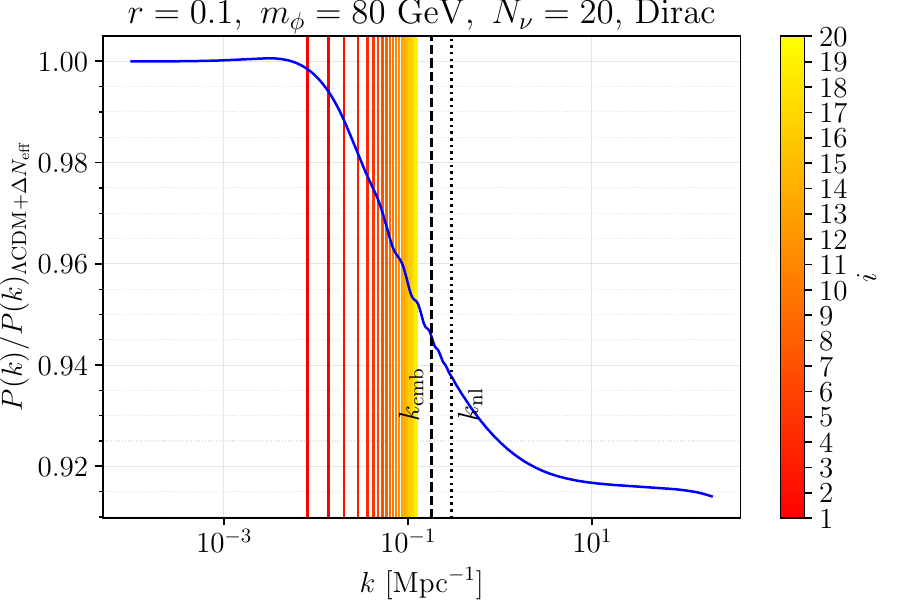}
  \qquad
  \includegraphics[width = 0.47\textwidth]{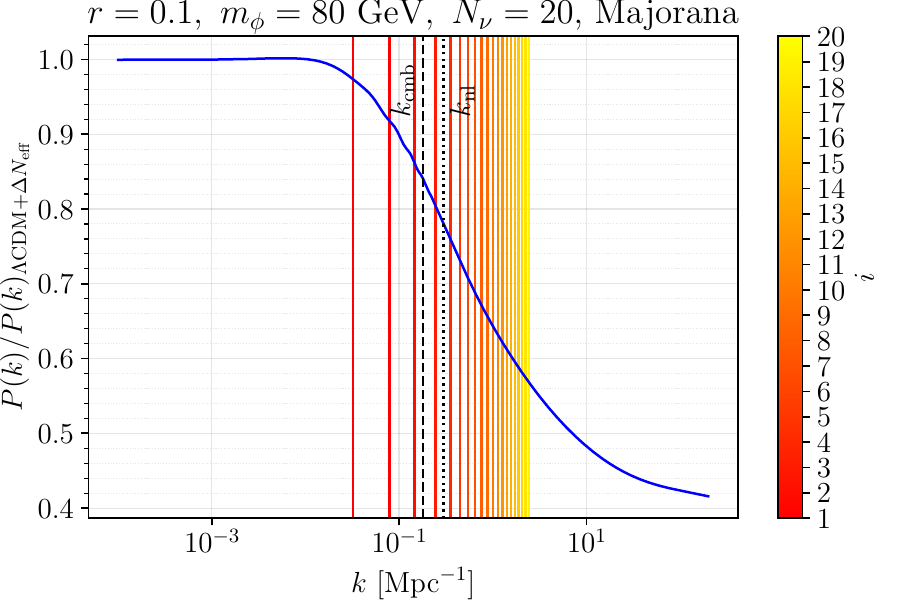}
  \caption{Ratio of the matter power spectrum of $N$naturalness to the matter power spectrum of $\Lambda$CDM$+\Delta N_{\rm eff}$ for Dirac neutrinos (left) and Majorana neutrinos (right).  The value of $\Delta N_{\rm eff}$ and $\omega_m$ in $\Lambda$CDM$+\Delta N_{\rm eff}$ are matched to the values at the $N$naturalness parameter point used.  The black dashed line corresponds to $k_{\rm cmb} = 0.18~{\rm Mpc}^{-1}$ which is the smallest scale measurable by Planck.  The black dotted line corresponds to $k_{\rm nl} = 0.3~{\rm Mpc}^{-1}$ which is the smallest scale that we compare against LSS data.  The colored lines indicate the wavenumber $k_{{\rm nr},i}$ at which the neutrinos from sector $i$ become non-relativistic.}
  \label{fig:k_nr}
\end{figure}

Figure~\ref{fig:k_nr} shows the ratio of the matter power spectrum in $N$naturalness to the matter power spectrum in $\Lambda$CDM as a function of wavenumber $k$.  The $N$naturalness parameters used are $r=0.1$, $m_\phi = 80~{\rm GeV}$, and $N_\nu=20$ and for the $\Lambda$CDM case the values of $\Delta N_{\rm eff}$ and the total matter density $\omega_m$ are adjusted to match the values in the $N$naturalness case.  The neutrinos of the SM are taken to be triply-degenerate with a sum of masses of $0.12~{\rm eV}$.  As we will show, this benchmark point is allowed at $1\sigma$ with Planck data when only the effects of $N$-photons are taken into account, however, it is nearly excluded at $2\sigma$ when effects of $N$-neutrinos are added. Thus, for this parameter point the $N$-neutrinos play a significant role in determining the total constraint.  The left plot shows the case of Dirac neutrinos and the right plot shows the case of Majorana neutrinos.  In both cases, there is a suppression of the matter power spectrum at small scales, but the scale at which the suppression starts occurring and the overall amount of suppression differs.

In Fig.~\ref{fig:k_nr}, two reference wavenumbers of $k_{\rm cmb}$ and $k_{\rm nl}$ are shown.  The first is $k_{\rm cmb} = 0.18~{\rm Mpc}^{-1}$ which is the approximate wavenumber up to which the Planck CMB measurement has sensitivity.  The second is $k_{\rm nl}$ which is where non-linear corrections to the matter power spectrum become significant.  This is relevant when fitting to LSS data from KV450 (see Sec.~\ref{sec:mcmc}) where only modes with $k<k_{\rm nl}$ are used in the fit.

To help understand the behavior of the matter power spectrum suppression we indicate the values of wavenumber, $k_{{\rm nr}, i}$, where the neutrinos from sector $i$ become non-relativistic, by vertical colored lines.  The wavenumber where a species of neutrino becomes non-relativistic can be approximated by the inverse of the conformal time that corresponds to when the temperature of the neutrino is equal to its mass
\begin{equation}
k_{{\rm nr},i} = \frac{1}{\eta_{{\rm nr},i}},
\quad\quad\quad
\text{where } T_{\nu,i}(\eta_{{\rm nr},i}) = m_{\nu,i}.
\end{equation}
This relation is only approximate since the high energy tail of the neutrino distribution is still relativistic when $T_\nu(\eta_{{\rm nr},i}) = m_{\nu,i}$. This gives, however, a fairly correct estimate of $k_{{\rm nr},i}$ when compared with numerical codes, which will be shown later.\footnote{Despite the fact that each species of neutrinos becomes non-relativistic at a different wavenumber because this process is not instantaneous, one does not resolve individual steps in the matter power spectrum ratio from each neutrino species.}  The matter power spectrum itself, along with the effects on the CMB, is calculated exactly by solving the full distribution function of the neutrinos.

Small-scale perturbations enter the horizon when many of the $N$-neutrinos are relativistic.  For such perturbations relativistic neutrinos do not contribute to the total matter fluctuations and, in fact, hinder matter clustering by contributing to radiation pressure.  Both of these effects slow down the total matter clustering and suppress the power spectrum~\cite{Lesgourgues:2006nd}.

Perturbations at larger scales enter the horizon later when the neutrinos of more sectors are non-relativistic.  These $N$-neutrinos act as cold dark matter and restore the matter power spectrum ratio with $\Lambda$CDM to unity.

The difference between the Dirac and Majorana cases stems from the difference in mass scales.  Eqs.~\eqref{eq:dirac} and~\eqref{eq:majorana} show the different dependence of $m_{\nu,i}$ on the VEV of its sector.  In terms of the sector index $i$, these equations are
\begin{equation}  \label{eq:nu_mass}
m_{\nu,i} = \begin{cases}
m_{\nu,{\rm SM}}\left(\frac{v_i}{v_{\rm SM}}\right) 
= m_{\nu,{\rm SM}} \sqrt{\frac{2i+r}{r}}
\quad\quad\quad &({\rm Dirac}), \\
m_{\nu,{\rm SM}} \left(\frac{v_i}{v_{\rm SM}}\right)^2
= m_{\nu,{\rm SM}} \left(\frac{2i+r}{r}\right)
\quad\quad\quad &({\rm Majorana}).
\end{cases}
\end{equation}
Therefore, the mass of Dirac neutrinos grows as $\sqrt{i}$ while the mass of Majorana neutrinos grows faster, according to $i$.  This is evident in Fig.~\ref{fig:k_nr} where the Dirac neutrinos become non-relativistic at comparatively lower wavenumbers than Majorana neutrinos.

In fact, the scaling of the wavenumber $k_{{\rm nr},i}$ where the neutrinos of sector $i$ become non-relativistic can be derived using Eq.~\eqref{eq:nu_mass} along with the relation of the temperature of sector $i$ relative to the temperature of the SM bath $T_i \approx i^{-1/4} T_{\rm SM}$ and the assumption that energy density of the Universe is dominated by the SM sector.  This lets us relate the temperature of a neutrino with the corresponding temperature of the SM bath and in turn with its conformal time.  This leads to
\begin{equation} \label{eq:k_f_scaling}
k_{{\rm nr},i} \approx k_{{\rm nr},{\rm SM}} \times \begin{cases}
i^{3/4}\quad\quad\quad &({\rm Dirac}),\\
i^{5/4}\quad\quad\quad &({\rm Majorana}).
\end{cases}
\end{equation}
The temperature relation between a sector $i$ and the SM also lets us estimate the scaling of the non-relativistic energy density in neutrinos
\begin{equation}
\label{eq:omega_nu_scaling}
\Omega_{\nu,i} = \Omega_{\nu,{\rm SM}} \times
\begin{cases}
\left(\frac{v_i}{v_{\rm SM}}\right)
\left(\frac{T_{\nu,i}}{T_{\nu,{\rm SM}}}\right)^3
\sim i^{-1/4}\quad\quad\quad &({\rm Dirac}), \\
\left(\frac{v_i}{v_{\rm SM}} \right)^2
\left(\frac{T_{\nu,i}}{T_{\nu,{\rm SM}}}\right)^3
\sim i^{1/4}\quad\quad\quad &({\rm Majorana}).
\end{cases}
\end{equation}
For Dirac neutrinos, the neutrinos from each successive sector have a decreasing non-relativistic energy density, but the overall non-relativistic neutrino energy density from all sectors still grows.  For the Majorana case, each successive sector itself already has an increasing non-relativistic energy density which leads to a sizable energy density and generally tighter constraints than for Dirac neutrinos.

This difference in energy density leads to a significantly larger suppression in the matter power spectrum at small scales for Majorana neutrinos than for Dirac neutrinos.  The amount of suppression can be estimated as~\cite{Hu:1997mj,Lesgourgues:2006nd}
\begin{equation}
\label{eq:8fnu}
\frac{\Delta P(k)}{P(k)} \approx -8f_\nu,
\quad\quad\quad
\text{for } k\gg k_{\rm nr},
\end{equation}
for small $f_\nu$, where $f_\nu$ is the ratio of the energy density of non-relativistic neutrinos and the total matter density
\begin{equation}\label{eq:fnuana}
f_\nu = \frac{\omega_\nu}{\omega_m}.
\end{equation}
When the energy density of non-relativistic neutrinos is large, at smaller scales these neutrinos are relativistic therefore energy density that would otherwise act as cold dark matter, now acts as radiation which does not cluster and contributes to radiation pressure at early times.

While Eq.~\eqref{eq:fnuana} provides a rough estimate of power spectrum suppression, a precise calculation is more complex. This complexity arises because some Dirac neutrinos remain relativistic even at the start of the matter-dominated era. Therefore, the matter power spectrum calculated numerically using \texttt{CLASS} is indispensable.

\begin{figure}
  \centering
  \includegraphics[width=0.47\textwidth]{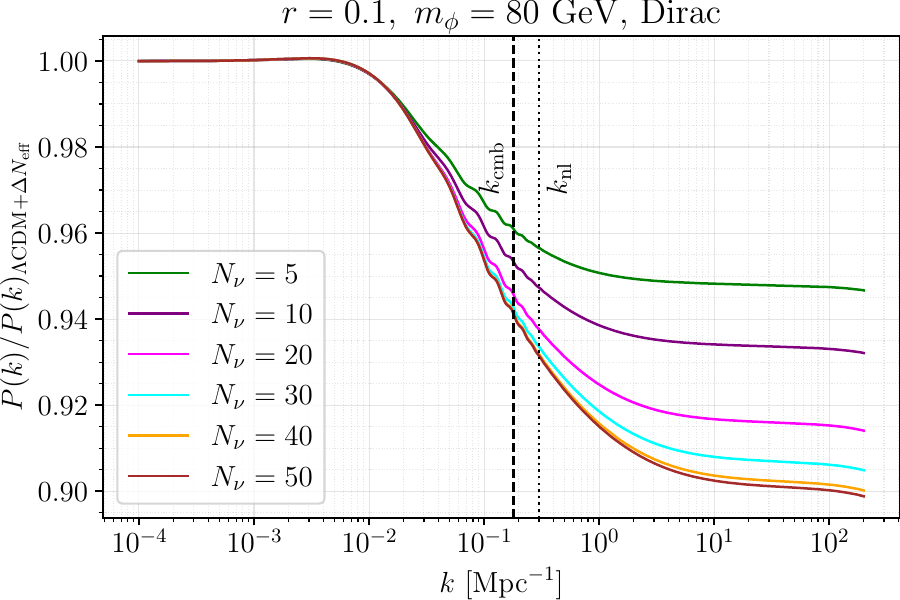}
  \qquad
  \includegraphics[width=0.47\textwidth]{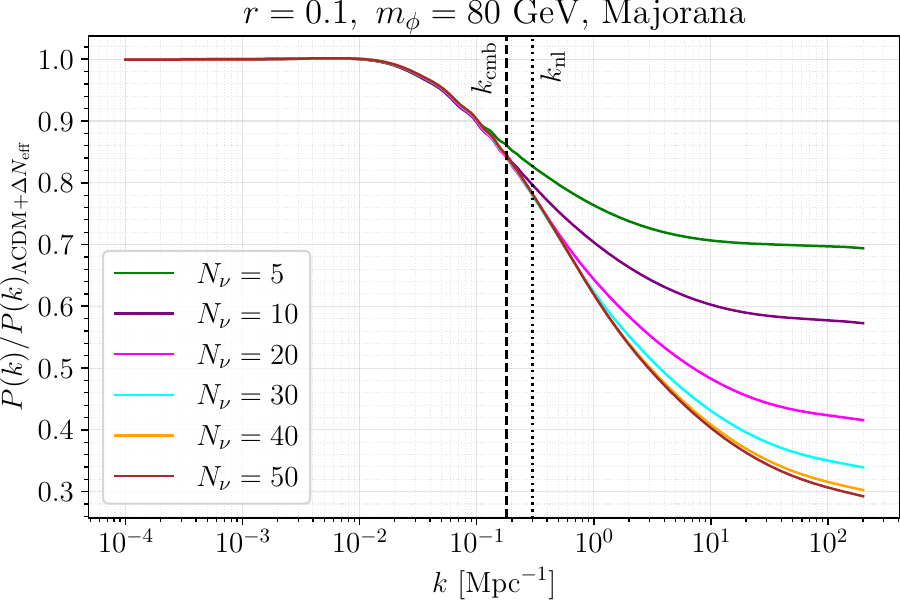}
  \caption{Ratio of the matter power spectrum for the benchmark $N$naturalness model with respect to the $\Lambda$CDM model. The values of $\omega_{cdm}$ and $\Delta N_{\rm eff}$ are adjusted so that the $N$naturalness and corresponding $\Lambda$CDM model have the same $\omega_{m}$ and $N_{\rm eff}$ values for Dirac neutrinos (left) and for Majorana neutrinos (right).}
  \label{fig:pk_ratio}
\end{figure}

In Fig.~\ref{fig:pk_ratio} we show the ratio of the matter power spectrum in $N$naturalness to the matter power spectrum in $\Lambda$CDM as a function of wavenumber $k$ for $N_\nu = \{5, 10, 20, 30, 40, 50 \}$.  Note that values of $\Delta N_{\rm eff}$ and $\omega_m$ for $\Lambda$CDM are adjusted to match the $N$naturalness case, separately for each value of $N_\nu$ shown.  As more sectors are included, the suppression grows larger, as expected from Eq.~\eqref{eq:omega_nu_scaling}.  This also shows that LSS and small-scale structure measurements will set useful constraints on the tower of neutrinos, especially for Majorana neutrinos.

For Majorana neutrinos, for the shown parameter space point, the suppression in the power spectrum converges at $k_{\rm cmb}$ and $k_{\rm nl}$ already for $N_\nu \gtrsim 10$.  Setting $N_\nu = 20$ captures the measurable effects of the $N$-neutrinos.  The reason for this behavior is that neutrinos from sectors $i\gtrsim 10$ are cold enough to be treated as cold dark matter and only neutrinos from sectors with $i\lesssim 10$ act as warm dark matter at that scale.

For Dirac neutrinos, the scenario is more subtle.  From Fig.~\ref{fig:k_nr} we see that $k_{{\rm nr}, i} < k_{\rm cmb}$ even for $i=20$ which would lead us to naively conclude that convergence should not occur near $N_\nu = 20$.  In contrast, the suppression of the power spectrum nearly does converge with $N_\nu = 20$ at $k_{\rm cmb}$ and $k_{\rm nl}$.  Perhaps $N_\nu = 30$ would be a more accurate value to use in our study, however, $N_\nu=20$ is already at the edge of computational tractability.  The near-convergence can be understood from Eq.~\eqref{eq:omega_nu_scaling} which shows that the energy density of sectors with larger $i$ noticeably decreases and as such cannot impact the matter power spectrum in a substantial way.

\begin{figure}
  \centering
  \includegraphics[width=0.47\textwidth]{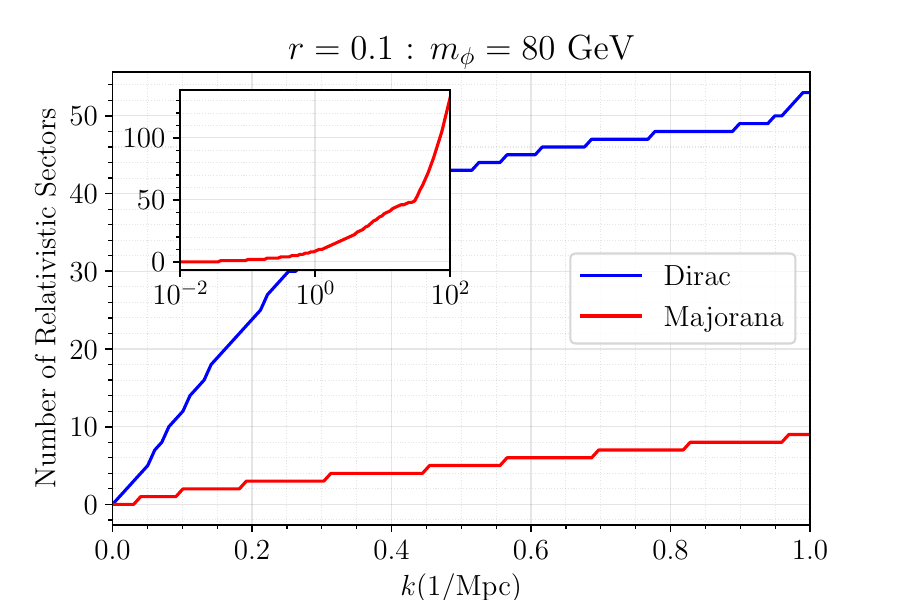}
  \qquad
  \includegraphics[width=0.47\textwidth]{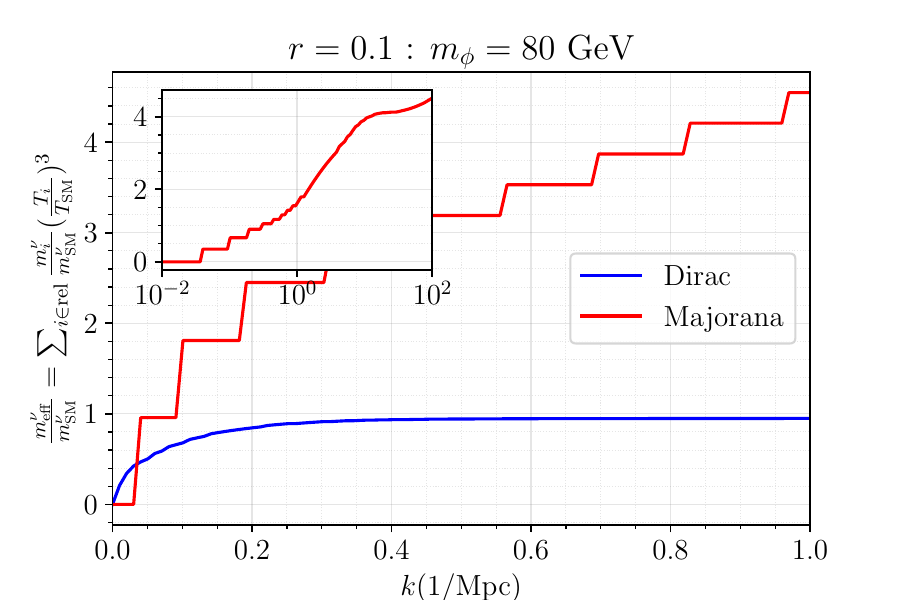}
  \caption{Number of relativistic sectors as a function of wavenumber (left).  The effective neutrino mass, as per Eq.~\eqref{eq:m_nu_eff}, is normalized by the SM neutrino mass as a function of wavenumber (right).  A neutrino is treated as relativistic if its temperature is above its mass.}
  \label{fig:relsector_w_k}
\end{figure}

Figure~\ref{fig:relsector_w_k} explores the dependence on $N_\nu$ further.  On the left, the number of sectors with relativistic neutrinos at the time of horizon re-entry for a given $k$-mode perturbation is shown.  For higher $ k$ modes, the number of sectors with relativistic neutrinos increases because those modes enter the horizon at earlier times.  For a given $k$-mode the number of sectors with relativistic neutrinos is much higher in the Dirac case than in the Majorana case because the neutrinos are much closer in mass in the Dirac case.

The right panel in Fig.~\ref{fig:relsector_w_k} utilizes the effective neutrino mass which is
\begin{equation} \label{eq:m_nu_eff}
m_{\nu,{\rm eff}} = \sum_{i\in {\rm rel}}  m_{\nu,i}
\left(\frac{T_{\nu,i}}{T_{\nu,{\rm SM}}}\right)^3,
\end{equation}
where $i \in {\rm rel}$ indicates that the summation is taken over the sectors with relativistic neutrinos at the horizon entry time of a given wavenumber.  The effective neutrino mass is proportional to the total matter density of neutrinos that remain relativistic at the horizon re-entry of a given $k$-mode.  The $k$-mode specifies how many sectors have relativistic neutrinos and contribute to the summation.

Figure~\ref{fig:relsector_w_k} (right) shows the ratio of the effective neutrino mass to the SM neutrino mass which determines the suppression in the power spectrum due to neutrinos becoming non-relativistic at a later time.  For Dirac neutrinos, the value of $m_{\nu, {\rm eff}}$ evaluated at a $k$-mode that is equivalent to $k_{{\rm nr}, i=20}$ (corresponding to $k \approx 0.18~{\rm Mpc}^{-1}$) compared to the value evaluated at $k_{{\rm nr}, i=50}$ (corresponding to $k \approx 1~{\rm Mpc}^{-1}$) differ by approximately 10\%.  Consequently, the difference in the matter power spectrum suppression, when increasing $N_\nu=20$ to $N_\nu=50$ is $\approx 0.5\%$.  Additionally, the value of $m_{\nu, {\rm eff}}$ plateaus at approximately $m_{\nu, {\rm eff}} = m_{\nu, {\rm SM}}$ and reaches the plateau around $k \approx 0.6~{\rm Mpc}^{-1}$.  For Majorana neutrinos, there is no such plateau over the range plotted.  The step-like behavior for Majorana neutrinos results from our implementation of the relativistic criteria of the temperature being greater than the mass.  The value of $m_{\nu,{\rm eff}}$ is higher for Majorana neutrinos than Dirac, despite the fact that there are more sectors of neutrinos included in the sum for Dirac neutrinos as shown in Fig.~\ref{fig:relsector_w_k} (left), because the mass of the neutrinos in each sector is larger for Majorana than for the corresponding Dirac case.

\begin{figure} 
  \centering
  \includegraphics[width=0.47\linewidth]{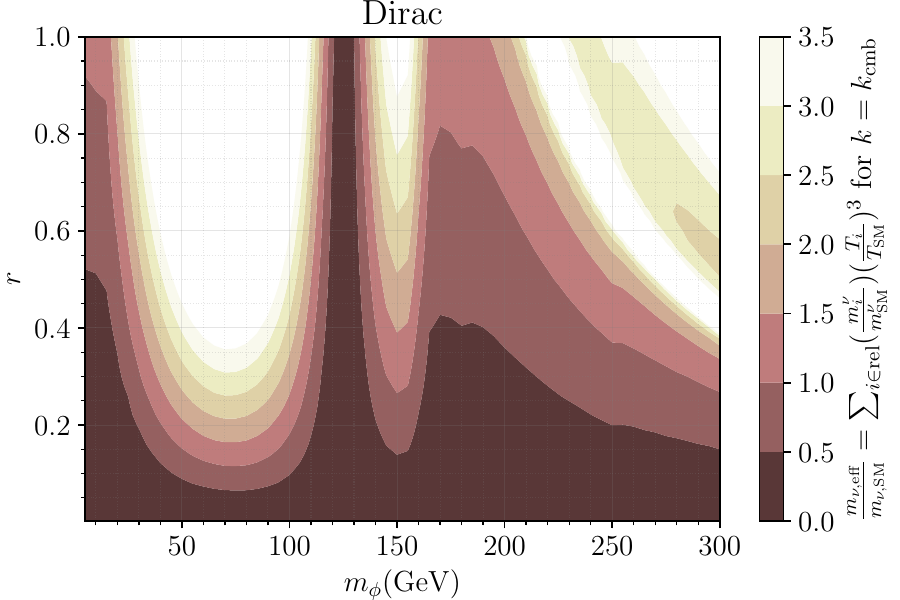}
  \qquad
  \includegraphics[width=0.47\linewidth]{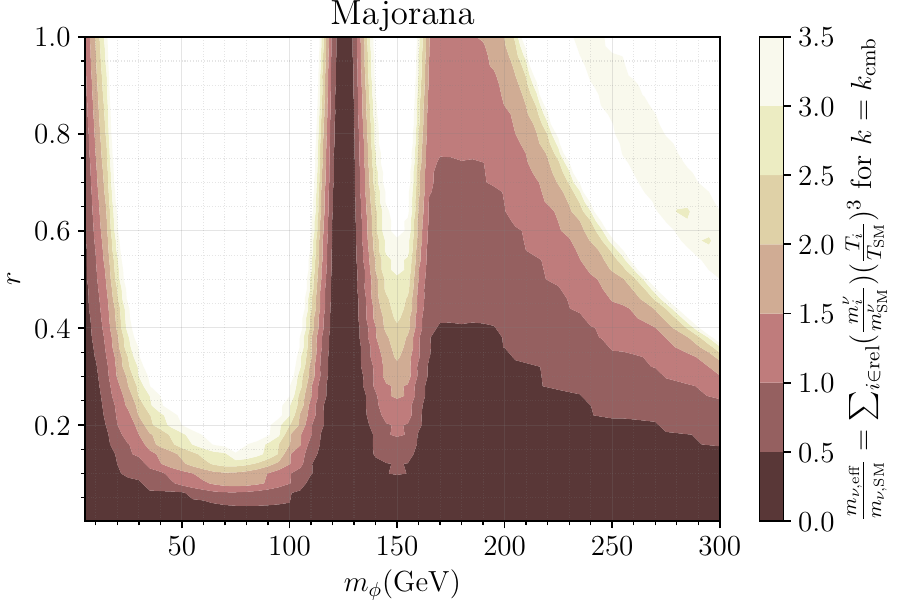} 
  \caption{The effective neutrino mass, as per Eq.~\eqref{eq:m_nu_eff}, normalized by the SM neutrino mass for Dirac neutrinos (left) and for Majorana neutrinos (right) evaluated at $k=k_{\rm cmb}$.}
  \label{fig:2d_nu}
\end{figure}

Finally, we show the ratio of the effective neutrino mass to the SM neutrino masses over the $N$naturalness parameter space of $r$ and $m_\phi$ in Fig.~\ref{fig:2d_nu}.  In order to distill the matter power spectrum suppression to a single value we take the value at $k=k_{\rm cmb}$.  We see that a suppression of the matter power spectrum generically occurs over the full parameter range.  Additionally, the general trend roughly follows the behavior of $\Delta N_{\rm eff}^\gamma$ in Fig.~\ref{fig:dneff_gamma} since both quantities grow with increasing temperatures of the additional sectors.  In detail, however, the behavior of $\Delta N_{\rm eff}^\gamma$ and the impact of $N$-neutrinos on the matter power spectrum is different because each has a different dependence on the temperature profile among the additional sectors.  The Dirac and Majorana cases have similar profiles because at $k_{\rm cmb}$ the values of $m_{\nu,{\rm eff}}$ are within a factor of 2, as seen in Fig.~\ref{fig:relsector_w_k}.  At a larger $k$, however, the equivalent profile of Fig.~\ref{fig:2d_nu}, will be much more constraining for the Majorana case.

\subsection{Effective Relativistic Degrees of Freedom}

When the temperature of a species of neutrino is well above its mass the neutrino behaves as radiation and contributes to $N_{\rm eff}$.\footnote{Since in the SM-like sectors the ordering of cosmological events (and in particular the fact that neutrino decoupling occurs before electron-positron annihilation~\cite{Choi:2018gho}) is the same as in the SM, the temperature ratio between the photons and the neutrinos is the same in SM-like sectors as it is in the SM.  This is not the case in the exotic sectors~\cite{Batell:2023wdb}.  While neutrinos from $i>0$ sectors will always contribute to $g_{*S}$ as they become non-relativistic their contribution to $g_*$ and consequently to $\Delta N_{\rm eff}$ will vanish~\cite{Batell:2023wdb}.}  The contribution characterized by $\Delta N_{\rm eff}$ is given by
\begin{equation} \label{eq:dneff_nu}
\Delta N_{\rm eff}^\nu = \sum_{i \in {\rm rel}} 3 \left( \frac{T_{\nu,i}}{T_{\nu,{\rm SM}}}\right)^4,
\end{equation}
where the $3$ is the number of neutrino species within a sector and the summation runs over the sectors where the neutrinos are relativistic.  Since the number of relativistic species of neutrinos is temperature-dependent and hence scale-dependent, the value of $\Delta N_{\rm eff}^\nu$ correspondingly depends on $k$ or equivalently, on redshift.\footnote{
Eq.~\eqref{eq:dneff_nu} neglects the effects of the small energy transfer between neutrinos and the photons after electron-positron annihilation (which is responsible for the SM value of $N_{\rm eff} = 3.044$~\cite{Gariazzo:2019gyi,Bennett:2020zkv}). The deviation from $N_\nu = 3$ for individual sectors will also be different for different sectors. However, this correction is fairly small in the context of additional contributions to $\Delta N_{\rm eff}$.}

\begin{figure}
  \centering
  \includegraphics[width=0.47\textwidth]{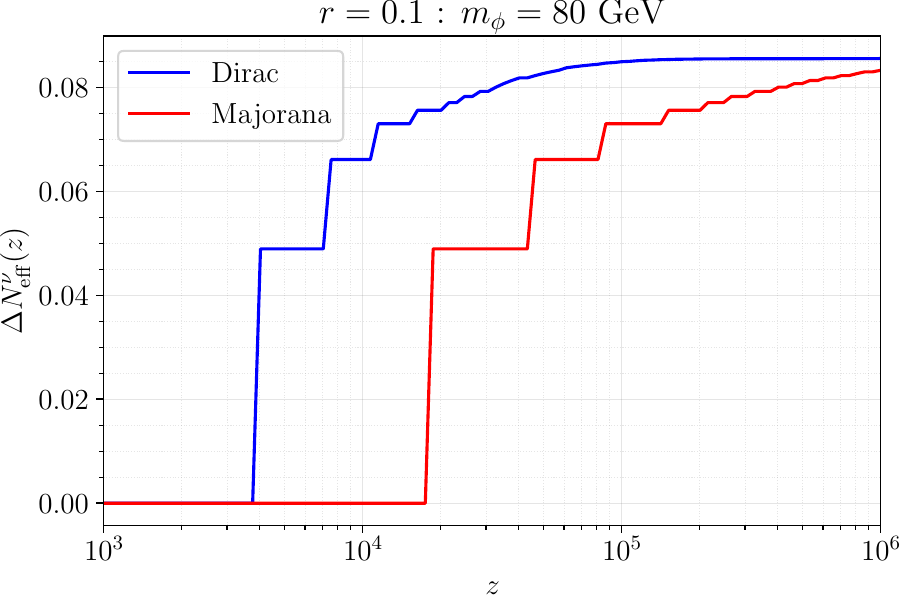}
  \caption{Scale-dependent contribution to $N_{\rm eff}$ from neutrinos as a function of redshift, $\Delta N_{\rm eff}^\nu(z)$.  Neutrinos contribute to $\Delta N_{\rm eff}^\nu(z)$ when they become relativistic and are considered to be relativistic when their temperature is larger than their mass.}
  \label{fig:neff_nu}
\end{figure}

At earlier times more sectors have neutrinos that are relativistic which causes the $\Delta N_{\rm eff}$ contribution from neutrinos, $\Delta N_{\rm eff}^\nu$, to increase with $z$.  Thus, models of $N$naturalness exhibit a scale-dependent $N_{\rm eff}$ from neutrinos, which is shown in Fig.~\ref{fig:neff_nu}.  At a fixed redshift, the Dirac case has more sectors with relativistic neutrinos than the Majorana case because Dirac neutrinos have smaller masses.  Correspondingly, $\Delta N_{\rm eff}^\nu$ is larger for Dirac neutrinos than for Majorana neutrinos.  At very high redshifts, both cases asymptote to a similar value of $\Delta_{\rm eff}^\nu$.

It is worth noting that at the last scattering surface, $z \approx 1100$, the $N$-neutrinos do not contribute to $N_{\rm eff}$.  In other words, the lightest of the $N$-neutrinos, corresponding to the neutrinos from the $i=1$ SM-like sector, has become non-relativistic by last-scattering.  Therefore, $\Delta N_{\rm eff}^\nu$ does not take part in silk damping which depends on the value of $\Delta N_{\rm eff}$ at the last scattering. However, the scale-dependent nature of  $\Delta N_{\rm eff}^\nu$ does modify the expansion history before the last scattering.  We will show quantitatively that among the two effects of $N$-neutrinos, the suppression of the matter power spectrum and the contribution to $\Delta N_{\rm eff}$, the former is more significant in driving the constraints on the $N$naturalness parameters.

\subsection{Cosmic Microwave Background}

\begin{figure}
  \centering
  \includegraphics[width = 0.47\textwidth]{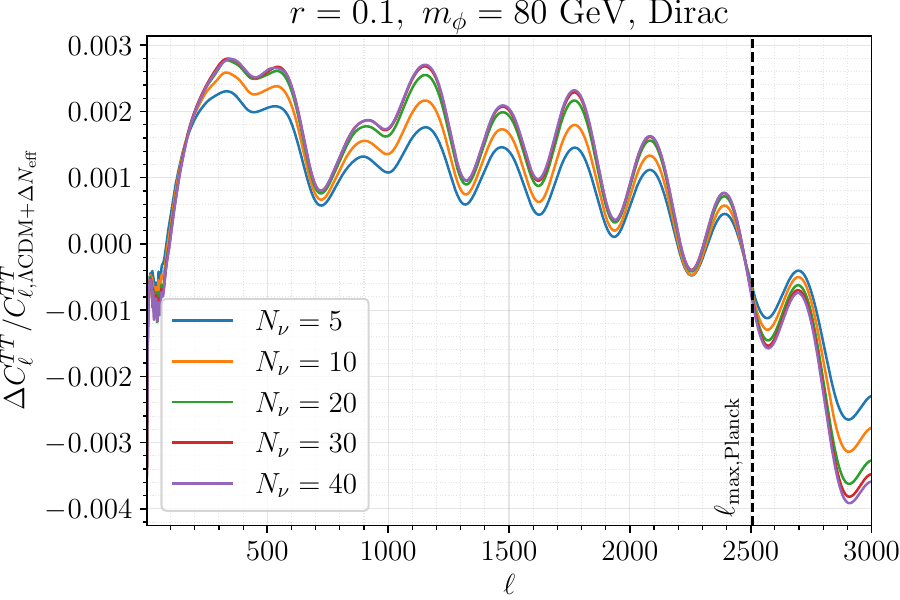}
  \qquad
  \includegraphics[width = 0.47\textwidth]{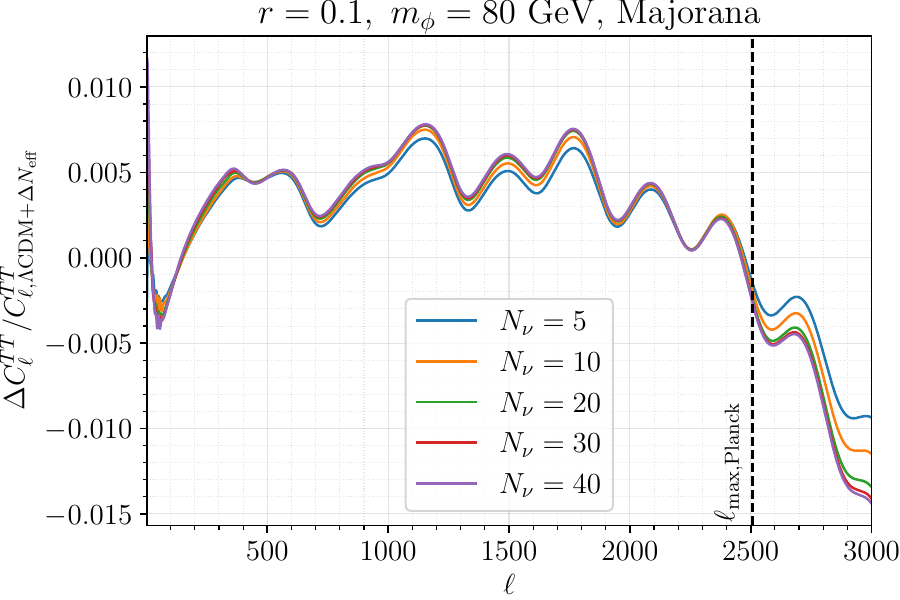}
  \caption{Fractional deviation of lensed TT power spectrum for the various $N$naturalness models relative to the $\Lambda$CDM model for Dirac neutrinos (left) and Majorana neutrinos (right). The values of $\omega_{cdm}$ and $N_{\rm eff}$ are adjusted so that the $N$naturalness and corresponding $\Lambda$CDM model has the same $\omega_{m}$ and $N_{\rm eff}$ values.  Both the $N$naturalness and $\Lambda$CDM spectra are computed at fixed $\theta_s$.  The $\Delta N_{\rm eff}$ value for a given $N$naturalness model is obtained when all neutrinos are relativistic.}
  \label{fig:cmb}
\end{figure}

In Fig.~\ref{fig:cmb}, we investigate the impact of $N$naturalness on the CMB by plotting the fractional change in the $C_\ell^{TT}$ coefficients $\Delta C_\ell^{TT}/C_\ell^{TT}$ as a function of multipole $\ell$.  As before, the change is with respect to $\Lambda$CDM with $\Delta N_{\rm eff}$ (referred to as $\Lambda$CDM$+\Delta N_{\rm eff}$).  For each value of $N_\nu$ the $\Lambda$CDM$+\Delta N_{\rm eff}$ parameters are set such that $\Delta N_{\rm eff}$ matches $N$naturalness at high redshifts where all $N$-neutrinos are relativistic and $\omega_m$ matches $N$naturalness today, which includes contributions from the $N$-neutrinos that are non-relativistic today.  The sound horizon angle, $\theta_s$, is set to be the same as well.  Thus, this comparison captures the redshift dependence of the $N$-neutrinos and how they differ from standard cold DM. 

The redshift-dependent contribution to $\Delta N_{\rm eff}$ from the $N$-neutrinos leads to $\ell$-dependent corrections to the power spectrum after fixing $\theta_s$.  Unlike massless radiation, the matter-energy density of $N$-neutrinos introduces redshift-dependent corrections to the Hubble expansion rate that can modify the phase of the photon's acoustic oscillations. The phase shift can cause oscillating ratios in Fig.~\ref{fig:cmb} between the $N$-neutrinos and free-streaming radiation scenarios. Similar to cosmological scenarios with interacting radiation~\cite{Bashinsky:2003tk,Baumann:2015rya,Pan:2016zla} or radiation with dark matter loading~\cite{Ghosh:2024wva}, the semi-relativistic behavior of $N$-neutrino propagation can also induce a phase shift in the acoustic peaks. A detailed analysis of this phase shift is left for future work. In addition to shifting the acoustic peaks, warm dark matter suppresses metric perturbations, leading to a reduction in the TT spectrum at high $\ell$. Both the suppression and phase shift are promising targets for future high-$\ell$ measurements, such as those from CMB-S4~\cite{CMB-S4:2016ple}.

Corrections to the CMB spectrum from using $N_\nu = 20$ are $\mathcal{O}(0.1\%)$ as can be seen in Fig.~\ref{fig:cmb} by comparing the $N_\nu$ line to the lines for larger $N_\nu$.  The correction is larger for Dirac neutrinos due to the presence of more light neutrino states compared to the Majorana case.  Additionally, the correction increases with $\ell$ which is expected based on the power spectrum dependence on $k$ in Fig.~\ref{fig:pk_ratio}.  The difference between $N_\nu = 20$ and $N_\nu=50$, while non-zero, is significantly below the current precision of CMB data.  For the CMB, these sectors act as cold dark matter to a good approximation and contribute primarily to the background total matter density.  Therefore, the effects of sectors with $i>20$ can largely be accounted for by varying the contribution of cold dark matter which we do in our study.\footnote{This fact, along with Eq.~\eqref{eq:omega_nu_scaling}, points out that as $N_\nu$ grows there is an eventual overclosure limit where the neutrino contribution to the matter density surpasses the measured cold dark matter density although this is not a relevant limit in our parameter space~\cite{Arkani-Hamed:2016rle}.}

As mentioned in Sec.~\ref{sec:mps}, motivated by computational accuracy and tractability, we set $N_\nu = 20$ in our full Monte Carlo study.  This turns out to be a sufficient choice also in light of the CMB.  As we have seen, for Majorana neutrinos $N_\nu = 20$ is nearly indistinguishable from using larger values of $N_\nu$ while for Dirac neutrinos we achieve a spectrum with an accuracy of $\approx 0.1\%$.

\begin{figure}
  \centering
  \includegraphics[width = 0.47\textwidth]{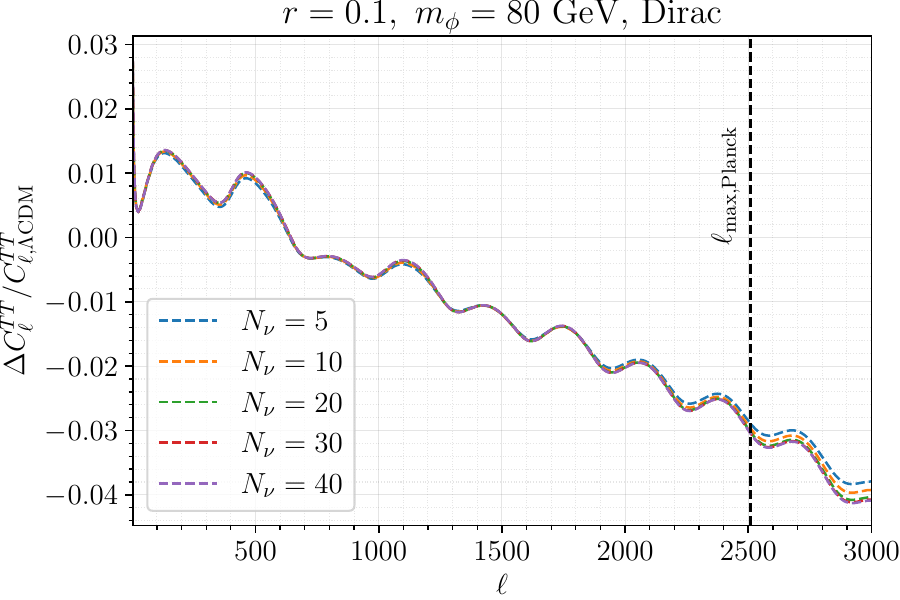}
  \qquad
  \includegraphics[width = 0.47\textwidth]{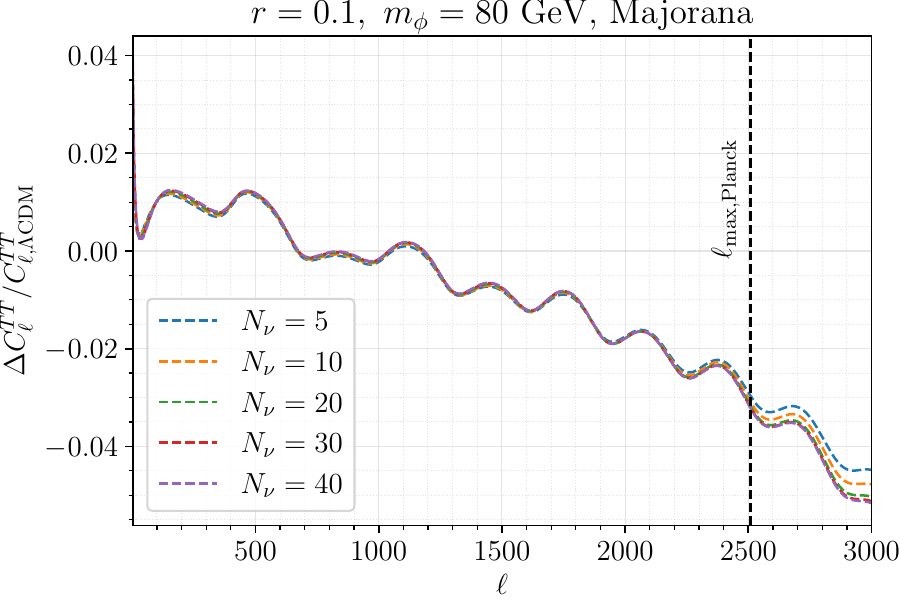}
  \caption{Fractional deviation of lensed TT power spectrum for the various $N$naturalness models relative to the $\Lambda$CDM model for Dirac neutrinos (left) and Majorana neutrinos (right). The value of $\omega_{cdm}$ only is adjusted so that the $N$naturalness and corresponding $\Lambda$CDM model has the same $\omega_{m}$. Unlike Fig.~\ref{fig:cmb} the $N_{\rm eff}$ of the $\Lambda$CDM curves are held fixed to the standard value.  Both the $N$naturalness and $\Lambda$CDM spectra are computed at fixed $\theta_s$. This comparison illustrates the effects of additional $N_{\rm eff}$, coming from both $N$-photon and $N$-neutrinos, which is the primary constraint for $N$naturalness models from CMB.}
  \label{fig:cmb_wo_neff}
\end{figure}

In Fig.~\ref{fig:cmb_wo_neff}, we plot the fractional change in the $C_\ell^{TT}$ coefficients $\Delta C_\ell^{TT}/C_\ell^{TT}$ as a function of multipole $\ell$.  This is similar to Fig.~\ref{fig:cmb}, except in Fig.~\ref{fig:cmb_wo_neff} we only adjust $\omega_{cdm}$ such that the $N$naturalness model and the corresponding $\Lambda$CDM model have the same $\omega_m$.  The value of $N_{\rm eff}$ for the $\Lambda$CDM models is not adjusted and is instead, fixed to $N_{\rm eff} = 3.046$.  The value of $N_{\rm eff}$ for the $N$naturalness model contains contributions from $N$-neutrinos and $N$-photons.  Thus for fixed $\theta_s$, Fig.~\ref{fig:cmb_wo_neff} illustrates the additional silk damping coming from extra contributions to $N_{\rm eff}$.  This is the primary constraint on the $N$naturalness model from the CMB.  These curves depend very mildly on $N_\nu$ because the additional $N_{\rm eff}$ is mostly from the $N$-photons, which do not depend on $N_\nu$, rather than from the $N$-neutrinos.

Finally, we briefly discuss large-scale structure measurements that are relevant for higher $k$-modes, starting first with the case of Dirac neutrinos.  As can be seen from Fig.~\ref{fig:pk_ratio} (left) achieving a sub-percent-level accuracy would require $N_\nu > 20$.  From Fig.~\ref{fig:k_nr} (left) we see that $k_{{\rm nr},i=20} < k_{\rm nl}$ which is why $N_\nu=20$ does not achieve sub-percent accuracy and instead underestimates the suppression of the matter power spectrum.  Regarding reconciling data with predictions, for data, we consider the KiDS$+$VIKING-450 (KV450) dataset~\cite{Hildebrandt:2018yau}, while for the computation we can only reliably compute the linear matter power spectrum.  Accordingly, we only include data for $k < k_{\rm nl}$ in our likelihood computation where we choose $k_{\rm nl} = 0.3~{\rm Mpc}^{-1}$.  Restricting the computation to this range leads to an error of $2-3\%$.

Moving onto the Majorana case, we see from Fig.~\ref{fig:k_nr} (right) that for $N_\nu = 20$, $k_{{\rm nr},i=20}>k_{\rm nl}$.  This means that if the neutrinos from additional sectors were included they would already be non-relativistic at the smallest scales used in the large-scale structure dataset and not impact structure formation.  This behavior is demonstrated clearly in Fig.~\ref{fig:pk_ratio} where $N_\nu=20$ is clearly sufficient.
 
The arguments we have made have used the single parameter point of $r=0.1$ and $m_\phi = 80~{\rm GeV}$.  This point resides close to the boundary of the allowed region according to $\Delta N_{\rm eff}$ alone (see Fig.~\ref{fig:dneff_gamma}).  Since $\Delta N_{\rm eff}$ is the leading constraint, this point will be as dissimilar from $\Lambda$CDM as is allowed by the data.  We expected, therefore, that the error estimates sourced by this parameter point are reliable in the relevant $N$naturalness parameter space.

\subsection{The Choice of Standard Model Neutrino Masses}

In this subsection, we discuss the impact of our choice of Standard Model neutrino masses.  As discussed in Sec.~\ref{sec.class} we use triply-generate SM neutrinos each with a mass of $m_{\nu, {\rm SM}} = 0.04~{\rm eV}$ or equivalently, a sum of $0.12~{\rm eV}$.  This is the heaviest possible sum of neutrino masses which means any other choice would be lighter and consequently there would be more sectors containing relativistic neutrinos at a given redshift.  This can be seen through Eq.~\eqref{eq:k_f_scaling} where $k_{\rm nr,SM}$ depends on the Standard Model neutrino mass.

The impact of more sectors containing relativistic neutrinos would be that $N_\nu$ would need to be larger to obtain the same levels of accuracy.  In principle, this is simple, but in practice, this would push against the well-known speed limitation of \texttt{CLASS} for computing with a large number of neutrino states~\cite{Diego_Blas_2011,Kamionkowski:2021njk}.

This issue could be approached by increasing the speed of {\tt CLASS} or of a {\tt CLASS}-like code.  In our work, we have implemented a few thoroughly-checked approximations but even so, the current speed is not nearly efficient enough to perform calculations for $N_\nu = \mathcal{O}(100)$.  Recent studies about the faster computation of perturbations of massive neutrinos without solving the Boltzmann equation hierarchy may be relevant for this case~\cite{Kamionkowski:2021njk,Ji:2022iji}.  An effective fluid formalism may be able to capture the effects of multiple warm DM states, but such an investigation is beyond the scope of this work.

The bounds that are quoted in this work are applicable for $\sum m_{\nu, {\rm SM}} = 0.12~{\rm eV}$. Choosing a lighter set of SM neutrino masses can relax the bounds.  The amount of suppression from the $N$-neutrinos scales approximately proportionally to the neutrino masses in each sector. Therefore, reducing $\sum m_{\nu, {\rm SM}} = 0.06~{\rm eV}$, for example, will reduce the suppression at the small scale roughly by a factor of 2 for the same number of total sectors. In that case, however, to reliably compute the power spectrum at the same $k$, one will need to include more sectors (since $k_{{\rm nr },i}(\sum m_{\nu, {\rm SM}} = 0.06~{\rm eV}) >k_{{\rm nr },i}(\sum m_{\nu, {\rm SM}} = 0.12~{\rm eV}) $) which in turn will enhance the suppression. 
Thus, Lyman-$\alpha$ bounds will change slightly but still apply, as the SM neutrino masses are changed.  The case of Majorana neutrinos, in particular, will still face strong constraints from small-scale probes.

\section{$N$-Neutrinos vs a Single Effective Species of Neutrinos}
\label{sec:N-nu}

The gradual suppression of the power spectrum from the $N$-neutrinos is unique.  In this section, we investigate the qualitative difference between matter power spectrum suppression effects between a single species of neutrino (``$1\nu$''), with properties tuned to match features of a tower of neutrinos, and a true tower of neutrinos (``$N\nu$''). 

The effects of $1\nu$ on the matter power spectrum can be quantified via three background quantities, $\Delta N_{\rm eff}$, $\omega_\nu^{\rm nr}$, and $\langle v_s \rangle$, which are not all independent for $1\nu$.  Here $\Delta N_{\rm eff}$ quantifies the relativistic energy density of the neutrinos, $\omega_\nu^{\rm nr}$ quantifies the non-relativistic energy density of the neutrinos, and $\langle v_s \rangle$ is the average velocity which can be related to the scale, $k_{\rm nr}$.  The average velocity for $1\nu$ is defined as
\begin{equation} \label{eq:vs}
\langle v_s \rangle =  \dfrac{\int p^3 dp f(p)}{ m\int p^2 dp f(p) } \approx 5.6 \times 10^{-6} \dfrac{\Delta N_{\rm eff}}{\omega_\nu^{\rm nr}}.
\end{equation}
It has been established that if two $1\nu$ scenarios have different $m_\nu$ values and phase space distributions, as long as their values of $(\Delta N_{\rm eff},\omega_\nu^{\rm nr},\langle v_s \rangle)$ are identical, the resulting matter power spectra will be very similar to each other~\cite{Cuoco:2005qr,Acero:2008rh}.\footnote{It follows from this fact that if the three neutrinos within a sector have masses that are similar to each other, the computed matter power spectrum is nearly the same as setting all three neutrinos to the same average mass.  This justifies the degenerate-mass simplification we make in this work.}

The signature for $N\nu$ is more intricate than the signature of $1\nu$.  It is straightforward to define $\Delta N_{\rm eff}$ and $\omega_\nu^{\rm nr}$ for $N\nu$, which denote the total relativistic and non-relativistic energy densities, respectively, for all neutrinos.  However, a generalization of  Eq.~\eqref{eq:vs} in the following manner:
\begin{equation} \label{eq:eff_vs}
\langle v_s \rangle_{\rm eff} =  \dfrac{\sum_i\int p^3 dp f_i(p)}{\sum_i m_i\int p^2 dp f_i(p) },
\end{equation} 
fails to capture the full essence of the suppression from the $N$-neutrinos.  As seen in Fig.~\ref{fig:k_nr} the neutrinos from a sector $i$ have their own corresponding scale $k_{{\rm nr},i}$ compared to a single scale $k_{\rm nr}$ for $1\nu$. 
The quantity $\langle v_s \rangle_{\rm eff}$ corresponds to a weighted average of different $k_{{\rm nr},i}$ values which is not a physical suppression scale observed in the power spectrum. 

\begin{figure}
  \centering
  \includegraphics[width=0.47\linewidth]{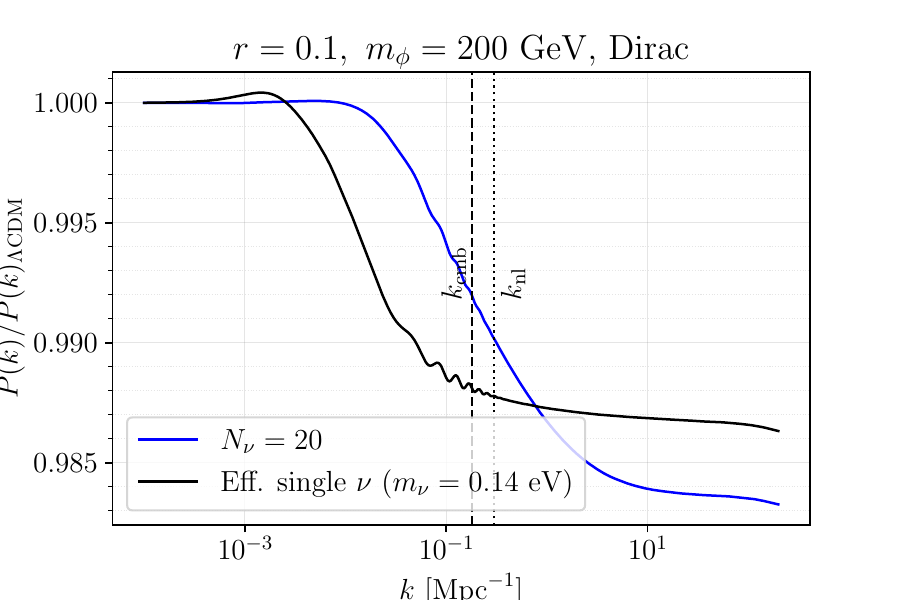}
  \qquad
  \includegraphics[width=0.47\linewidth]{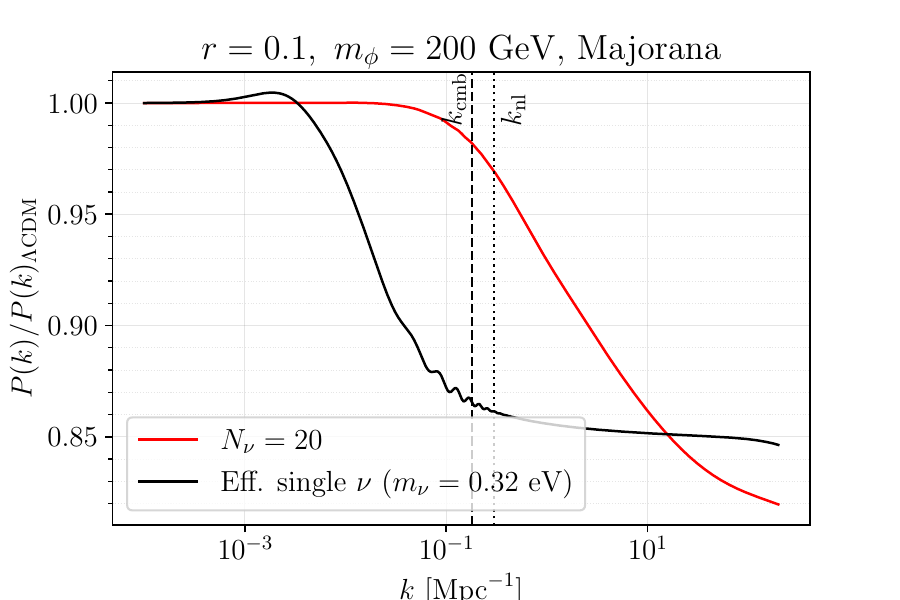}
  \caption{Ratio of the matter power spectrum of $N$naturalness to the matter power spectrum of $\Lambda$CDM$+\Delta N_{\rm eff}$ for Dirac neutrinos (left) and Majorana neutrinos (right), with fixed Hubble expansion $h$ today. The value of $\Delta N_{\rm eff}$ and $\omega_m$ in $\Lambda$CDM$+\Delta N_{\rm eff}$ are matched to the values at the $N$naturalness parameter point used.  The numerators of the blue and red lines correspond to the $N\nu$ scenario of the $N$naturalness model with $N_\nu=20$ while the numerators of the black lines correspond to the $1\nu$ scenario of a single effective species of neutrino.  In the $1\nu$ case, the total mass and the temperature of the neutrino species are adjusted to match the values of $N_{\rm eff}$ and $\omega_m(z=0)$ of the $N\nu$ case.}
  \label{fig:eff_nu}
\end{figure}

In Fig.~\ref{fig:eff_nu}, we show the ratio of the matter power spectrum between $N$naturalness and $\Lambda$CDM$+\Delta N_{\rm eff}$.  The $\Lambda$CDM$+\Delta N_{\rm eff}$ spectra match the values of $\Delta N_{\rm eff}$ and $\omega_m$ to the associated $N$naturalness scenario.  The blue and red lines show the ratios for the $N\nu$ scenario with $N_\nu=20$ and the black lines show the ratios for the $1\nu$ scenario.  In the $1\nu$ scenario the value of $\Delta N_{\rm eff}$ is chosen to match the $N\nu$ scenario at early times and the value of $\omega_m$ is chosen to match the $N\nu$ scenario at late times.  The difference between these two spectra, therefore, encodes only the effects of multiple free-streaming scales of $N\nu$ vs. the single free-streaming scale of $1\nu$.

Regarding the magnitude of the matter power spectrum suppression, at small scales, both the $N\nu$ and $1\nu$ scenarios achieve a similar suppression because they have the same value of $\omega_m$.  The asymptotic suppression at large $k$, however, is not identical due to the fact that matter radiation equality changes between the $N\nu$ and $1\nu$ cases.  The suppression as a function of $k$ for $N\nu$ is more gradual compared to the $1\nu$ case.  This feature is a result of the range of $k_{{\rm nr}, i}$ values present in the $N\nu$ case (see Fig.~\ref{fig:k_nr}).

The scale at which the matter power spectrum suppression starts to become significant also differs between $N\nu$ and $1\nu$.  The suppression in the $1\nu$ case starts at larger scales than $N\nu$ which means the free-streaming scale of $1\nu$ is larger than the free-streaming scale of each sector in the $N\nu$ case.  A suppression that extends to lower $k$-modes is a characteristic feature of using Eq.~\eqref{eq:eff_vs} to match $N$-neutrinos where the mass of neutrinos in each sector increases and the temperature of neutrinos in each sector generally decreases.  Consequently, large-scale structure measurements will more strongly constrain the effective single neutrino case than the associated case with $N$-neutrinos.

This is directly demonstrated with the $N$naturalness parameter point used in Fig.~\ref{fig:eff_nu} which is allowed by the Planck dataset within $2\sigma$ both for Dirac and Majorana neutrinos.  The $1\nu$ case for the Majorana, on the other hand, is strongly excluded by Planck at $2\sigma$.  This can be seen by the value of the sizable effective neutrino mass of $m_{\nu,{\rm eff}} = 0.14~{\rm eV}$ for Dirac neutrinos and $m_{\nu,{\rm eff}} = 0.33~{\rm eV}$ for Majorana neutrinos, whereas the Planck $2\sigma$ upper limit is $m_{\nu,{\rm eff}} = 0.24~{\rm eV}$~\cite{Planck:2018vyg}.  We conclude that $N$-neutrinos are qualitatively different than a single warm DM species and more effectively evade constraints by essentially siphoning the matter power suppression to smaller scales.

\section{MCMC Analysis}
\label{sec:mcmc}

\begin{table}
  \centering
  \begin{tabular}{|c|c|}
  \hline
         Parameters & Range \\
         \hline
         $10^2\omega_b$ & $[1.0,  4.0]$ \\
         \hline
         $\omega_{c}$ & $[0.08, 0.16]$ \\
         \hline
         $H_0~[{\rm km/s/Mpc}]$ & $[55.0, 85.0]$\\
         \hline
         $\ln 10^{10}A_s$ & $[2.0,  4.0]$ \\
         \hline
         $n_s$ & $[0.8,  1.1]$ \\
         \hline
         $\tau_{\rm reio}$ & $[0.004, 0.25]$ \\
         \hline
         $r$ & $[0.0,  1.0]$ \\
         \hline
         $m_\phi~[{\rm GeV}]$ & $[0.0,300.0]$\\
         \hline
  \end{tabular}
  \caption{Prior ranges for $N$naturalness  parameters used for MCMC calculation.}
  \label{tab:prior_table}
\end{table}

In this section, we perform an MCMC analysis over the $N$naturalness parameter space using CMB and LSS datasets.  This parameter space includes the usual $\Lambda$CDM parameters, augmented by $r$ and $m_\phi$.  Table~\ref{tab:prior_table} lists these parameters and the prior ranges used for the MCMC scan.  We use the following datasets for the analysis:
\begin{itemize}

  \item \underline{Planck} : This dataset contains low-$\ell$ TT $(\ell < 30)$ data, low-$\ell$ EE $(\ell < 30)$ data, high-$\ell$ TTTEEE data, and lensing data~\cite{Planck:2019nip}. 

  \item \underline{BAO}: We use the Baryon Acoustic Oscillation (BAO) dataset which is comprised of the 6DF Galaxy survey, SDSS-DR7 MGS data, the BOSS measurement of the BAO scale, and $f\sigma_8$ from the DR12 galaxy sample~\cite{Beutler_2011,Ross:2014qpa,BOSS:2016wmc}
  
  \item \underline{KV450}: We use the cosmic shear dataset KV450 to constrain the amplitude of the matter power spectrum at linear scales~\cite{Hildebrandt:2018yau}.  Due to our inability to reliably compute matter power spectra at smaller scales, we restrict the theoretical matter power spectra computation to $k \leq 0.3 ~{\rm Mpc}^{-1}$.

\end{itemize}

The MCMC analysis is performed using \texttt{Montepython}~\cite{Audren:2012wb,Brinckmann:2018cvx}. We used the Metropolis-Hastings algorithm and the Gelman-Rubin convergence criteria $R-1 <0.03$ was satisfied~\cite{Gelman:1992zz}.  For plotting we used the python package \texttt{GetDist}~\cite{Lewis:2019xzd}.  We used minimal smoothing (smoothing by bin width) to highlight the features of the posteriors.

\begin{figure}
  \centering
  \includegraphics[width=0.47\textwidth]{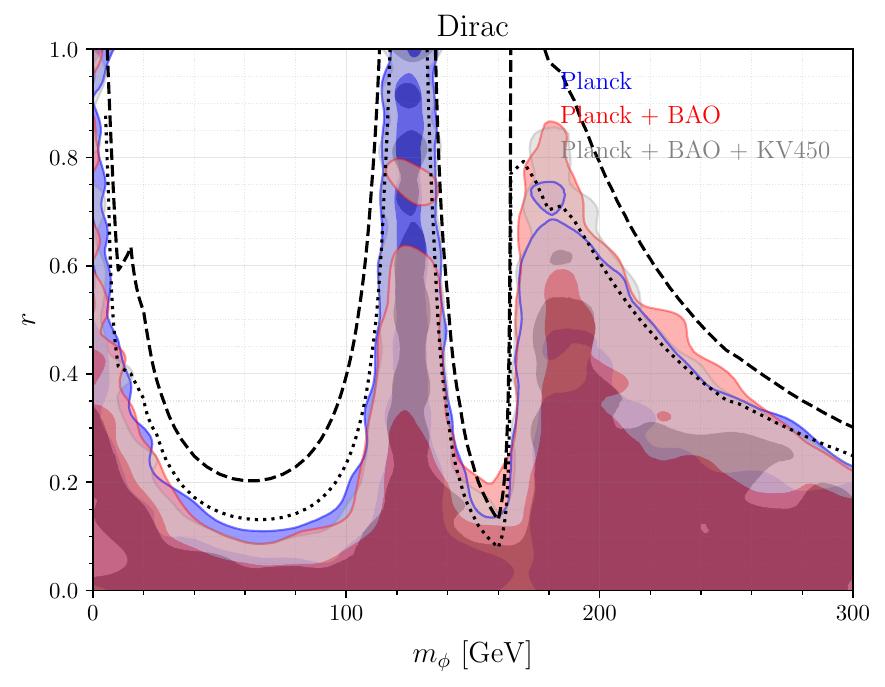}
  \qquad
  \includegraphics[width=0.47\textwidth]{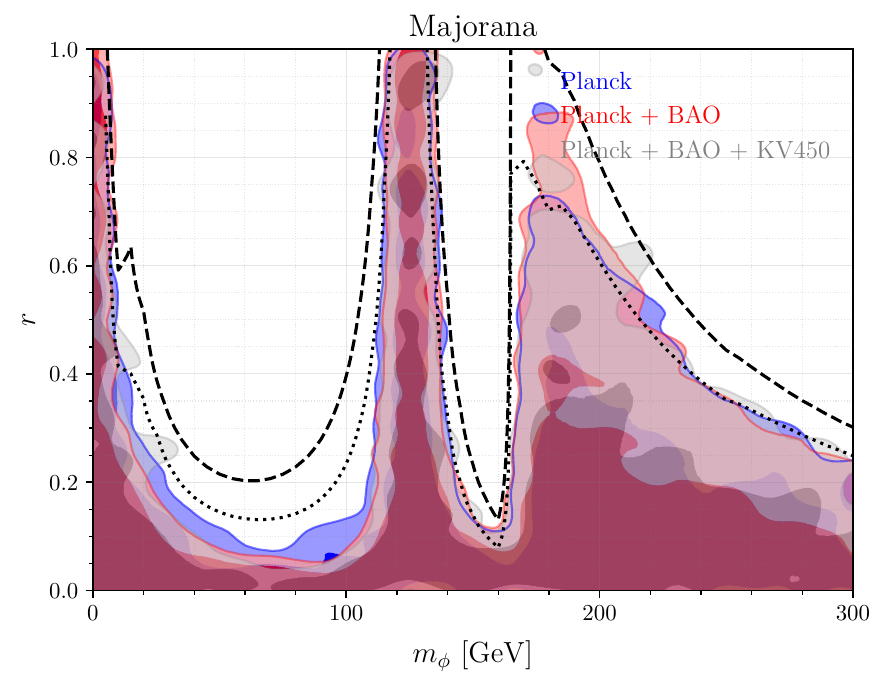}
  \caption{$2D$ posterior distributions of the $N$naturalness model over $r$ and $m_\phi$ for Dirac neutrinos (left) and Majorana neutrinos (right).  The black dotted (dashed) lines show the $1\sigma$ ($2\sigma$) Planck constraint excluding the $N$-neutrinos ({\it i.e.}, considering only the $N$-photons).  The blue, red, and gray regions show the constraints using data from Planck, Planck + BAO, and Planck + BAO + KV450, respectively.}
  \label{fig:2d_r_mphi}
\end{figure}

Figure~\ref{fig:2d_r_mphi} shows one of the primary results of our work, namely the marginalized constraints in the space of $r$ and $m_\phi$ from the datasets.  The black dotted and dashed contour lines show constraints solely from $\Delta N_{\rm eff}^\gamma$ at $1\sigma$ (corresponding to $\Delta N_{\rm eff}^\gamma = 0.168$) and $2\sigma$ (corresponding to $\Delta N_{\rm eff}^\gamma = 0.345$), respectively, from the Planck dataset.\footnote{To derive these limits we run $\Lambda$CDM+$\Delta N_{\rm eff}$ model with $m_\nu = 0.12~\mathrm{eV}$ and restrict $\Delta N_{\rm eff}$ to be above its SM value of 3.046.}  The blue, red, and gray contours show MCMC constraints with the datasets of Planck, Planck + BAO, and Planck + BAO + KV450, respectively, which includes the full impact of the $N$-neutrinos.

The results from Planck alone are quite similar for Dirac and Majorana neutrinos.  The similarity arises because the power spectrum suppression at $k_{\rm cmb} = 0.18~{\rm Mpc}^{-1}$, the scales down to which Planck is sensitive, is comparable with Dirac or with Majorana neutrinos.  This was discussed at length in Sec.~\ref{sec:mps} and most directly seen in Fig.~\ref{fig:2d_nu}.  In the Planck + BAO dataset, the constraints start to become tighter for the Majorana case compared to the Dirac case.  This is especially evident in the region where $m_\phi \gtrsim 160~{\rm GeV}$.  BAO data itself tightens constraints on neutrino mass~\cite{Planck:2013pxb} which translates to a tighter constraint on Majorana neutrinos whose $N$-neutrinos are more massive.  Including the weak lensing data from KV450 does not significantly improve the constraints for either neutrino scenario.  While using the KV450 data, we only include modes up to $k_{\rm nl} = 0.3~{\rm Mpc}^{-1}$, which is only slightly above $k_{\rm cmb}$. 

Overall, data from the CMB and from LSS does substantially reduce the available $N$naturalness parameter space.  Across the full parameter space, the Majorana neutrino case faces stronger constraints due to the larger suppression of the matter power spectrum.  Depending on the value of $m_\phi$ the corresponding tuning that is probed ranges from percent-level to upwards of $20\%$.

\begin{figure}
  \centering
  \includegraphics[width=0.8\textwidth]{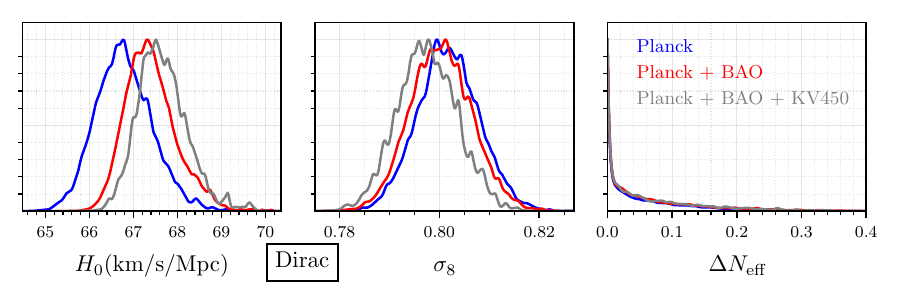}
  \includegraphics[width=0.8\textwidth]{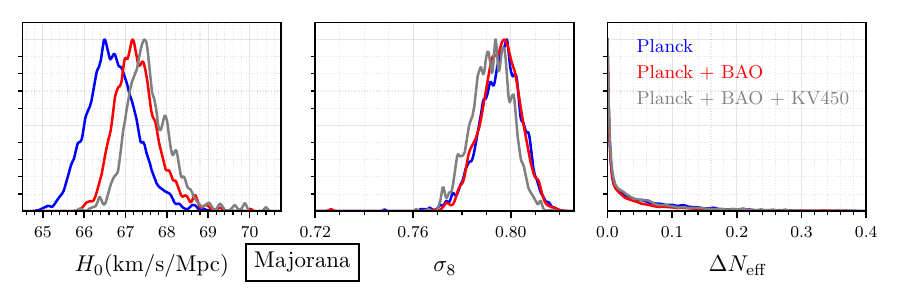}
  \caption{$1D$ posterior distributions of cosmological parameters of $H_0$, $\sigma_8$, and $\Delta N_{\rm eff}$ for Planck (blue), Planck + BAO (red), and Planck + BAO + KV450 (gray) for Dirac neutrinos (top) and Majorana neutrinos (bottom).}
  \label{fig:1d_pos_set}
\end{figure}

In Fig.~\ref{fig:1d_pos_set}, we show the $1D$ posterior distributions of several parameters of interest: $H_0$, $\sigma_8$, and $\Delta N_{\rm eff}$.  For all datasets, the $\Delta N_{\rm eff}$ distribution peaks towards zero but has a tail that does allow for non-zero values.\footnote{The value of $\Delta N_{\rm eff}$ output from {\tt CLASS} is taken in the ultra-relativistic limit of neutrinos meaning that all neutrinos -- even very massive species -- contribute to $\Delta N_{\rm eff}$.  In practice, for $N$naturalness this is not problematic because the quantity of $\Delta N_{\rm eff}$ receives almost its entire contribution from sectors with very low $i$ due to the decreasing temperature of successive sectors.}  For both neutrino cases, the value of $\sigma_8$ has the highest likelihood at smaller values than in $\Lambda$CDM because in $N$naturalness a fraction of the DM energy density is comprised of warm DM that suppresses matter perturbations~\cite{Planck:2018vyg}.  In the Majorana case, in particular, a significant fraction of DM is made up of $N$-neutrinos which is evidenced by the longer tail of $\sigma_8$ towards smaller values.  The KV450 dataset favors slightly lower values of $\sigma_8$, which can be achieved by reducing the CDM abundance. To maintain consistency with the expansion history at lower redshifts, this also results in a higher $H_0$.

\begin{figure}
  \centering 
  \includegraphics[width=0.47\textwidth]{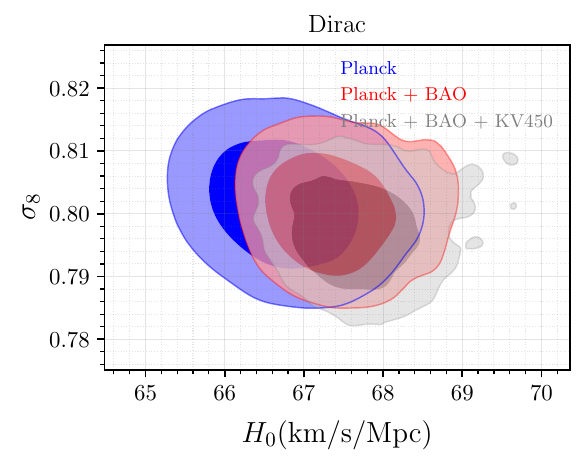}  
  \qquad
  \includegraphics[width=0.47\textwidth]{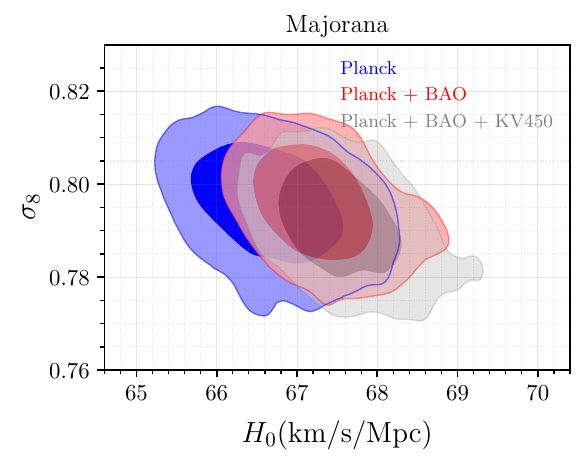} 
  \caption{$2D$ posterior distribution for $\sigma_8$ and $H_0$ for Planck (blue), Planck + BAO (red), and Planck + BAO + KV450 (gray).}
  \label{fig:2d_pos_set}
\end{figure}

The $2D$ posteriors for $H_0$ and $\sigma_8$ are shown in Fig.~\ref{fig:2d_pos_set}.  There is a slight negative correlation between $H_0$ and $\sigma_8$ which is more pronounced for Majorana neutrinos.  This correlation arises because each additional sector adds both a photon species and a neutrino species.  As more radiation is added, more warm DM is necessarily added.  This feature of $N$naturalness is highly interesting because, for standard massive neutrinos, the correlation between $H_0$ and $\sigma_8$ is positive~\cite{Planck:2013pxb}.  Simultaneously resolving the $H_0$ and $\sigma_8$ tensions requires a negative correlation.

\section{Estimated Constraint from Lyman-$\alpha$ }
\label{sec:lyalpha}

The suppression of the matter power spectrum at higher $k$-modes, as shown in Fig.~\ref{fig:k_nr}, indicates that small-scale structure measurements, particularly on scales smaller than those measured by Planck ($\approx 0.1~{\rm Mpc}^{-1}$), can effectively constrain models with a tower of warm DM states. In this section, we estimate how Lyman-$\alpha$ (Ly-$\alpha$) data can constrain the $N$-neutrinos.

The Ly-$\alpha$ likelihood we use is sensitive to the matter power spectrum at  $k \approx 1~ {\rm Mpc}^{-1}$.  In the $N$naturalness model, accurately describing the power spectrum suppression at $k \approx 1~ {\rm Mpc}^{-1}$ would require $N_\nu \gtrsim 40$ sectors of neutrinos.  In fact, the standard Ly-$\alpha$ likelihood available for use in MontePython~\cite{Brinckmann:2018cvx} requires computing the power spectrum to even higher wavenumbers of $k \approx 100~{\rm Mpc}^{-1}$.  Due to speed limitations of warm DM calculations in \texttt{CLASS} we are unable to include Ly-$\alpha$ data in the MCMC analysis.  To still gain insight, however, we can perform the following analysis instead. The dataset we use is:
\begin{itemize}

 \item \ul{Compressed eBOSS}: The eBOSS flux power spectrum can be compressed into two variables: the linear amplitude of the linear matter power spectrum $\Delta^2_{\rm lin} \equiv k_p^3 P_{\rm lin}(k_p,z_p)/(2\pi^2)$ and the tilt of the power spectrum $n_{\rm lin} \equiv d\ln P_{\rm lin}(k,z)/ dk|_{k_p,z_p}$ evaluated at redshift $z_p = 3$ and $k_p = 1.03h ~{\rm Mpc}^{-1} $~\cite{eBOSS:2018qyj,Goldstein:2023gnw,Rogers:2023upm,Bagherian:2024obh}. Following Ref.~\cite{Goldstein:2023gnw}, we use $\Delta^2_{\rm lin} = 0.310 \pm 0.020$ and $n_{\rm lin} = -2.340 \pm 0.006$ with correlation coefficient $0.512$ between them. We construct a Gaussian likelihood using these. 
 
\end{itemize}

We take the best-fit values from the MCMC analysis of $N$naturalness from Sec.~\ref{sec:mcmc} using the Planck+BAO dataset.  The subset of parameters corresponding to $\Lambda$CDM parameters are fixed to these best-fit values while the subset of parameters corresponding to $N$naturalness parameters are allowed to vary.  The deviations of the matter power amplitude and slope with respect to the $\Lambda$CDM benchmark were compared against the error bar from the compressed Ly-$\alpha$ likelihood to constrain the matter power spectrum.

In the analysis, the $N$naturalness parameters that are varied are $r$ and $m_\phi$.  Unlike the rest of the study, we fix the number of sectors of neutrinos to $N_\nu = 50$ for both the Dirac and Majorana neutrinos.  For the Majorana case, sectors with $N_\nu > 50$ become non-relativistic before horizon entry for $k \sim 1 ~{\rm Mpc}^{-1}$ modes, thus acting effectively as cold DM at the Ly-$\alpha$ scale.  In contrast, for the Dirac case, the effects of $N_\nu > 50$ are comparatively small, resulting in an $\mathcal{O}(1\%)$ correction to $P_{\rm lin}(k)$. Therefore, $N_\nu = 50$ serves as an optimal number of sectors to estimate the effects at the Ly-$\alpha$ scale.  Similar to the procedure in Sec.~\ref{sec:neutrinos}, for each $N$naturalness parameter point, we adjust the value of $\omega_{\rm cdm}$ to keep the total $\omega_{\rm m}$ the same after including the $N_\nu = 50$ sectors of neutrinos.  This ensures that the suppression and change in the slope of the matter power spectrum are solely due to the free-streaming effects of the $N$-neutrinos, while the background cosmology remains identical.

\begin{figure}
  \centering
  \includegraphics[width = 0.47\linewidth]{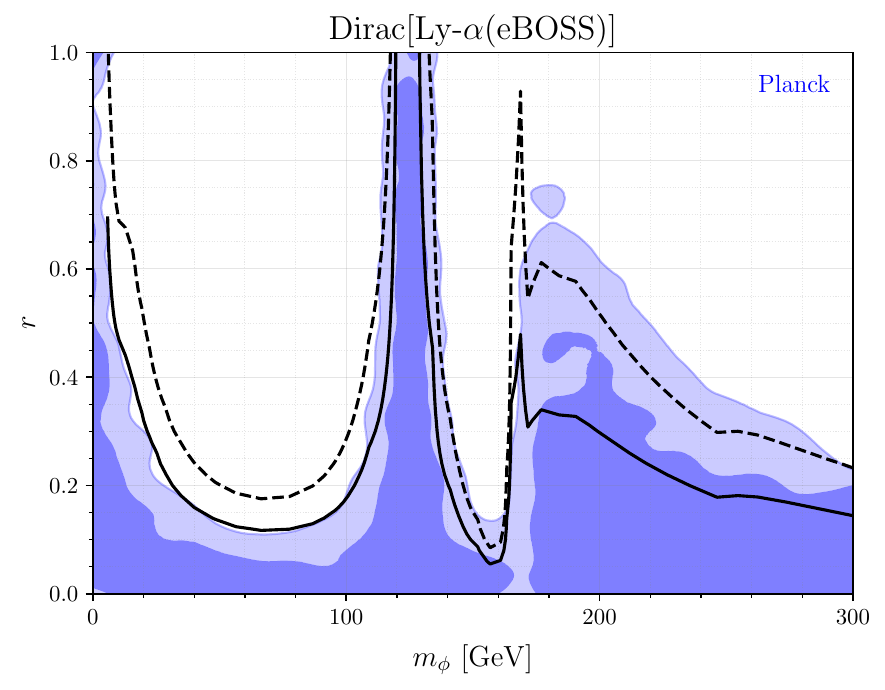}
  \qquad
  \includegraphics[width=0.47\linewidth]{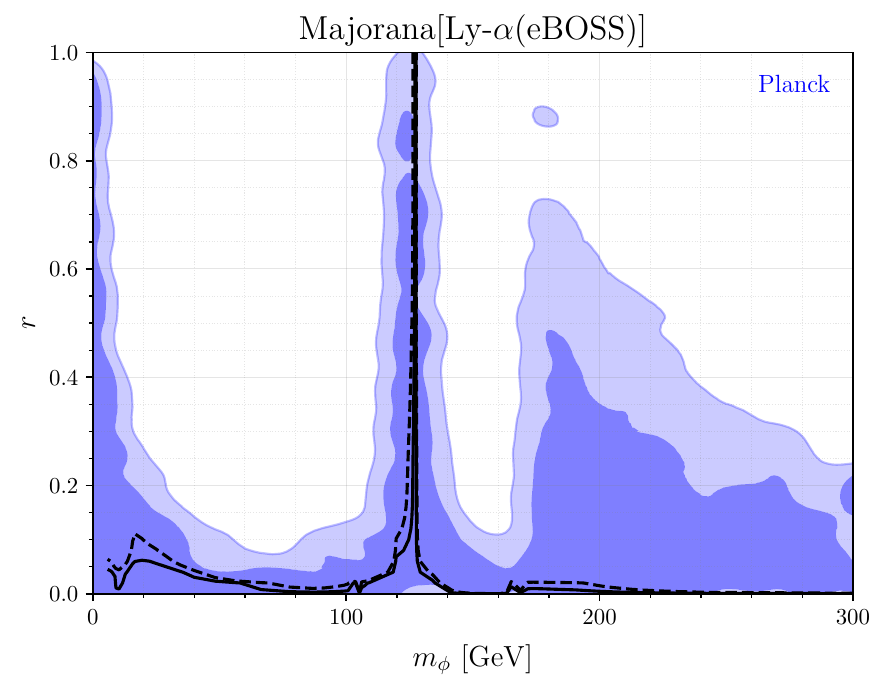}
  \caption{Estimated $1\sigma$ (solid black) and $2\sigma$ (dashed black) constraints on the $N$naturalness parameters from the compressed eBOSS  Ly-$\alpha$ likelihood alone for Dirac neutrinos (left) and for Majorana neutrinos (right).  MCMC results from Planck alone are shown at $1\sigma$ (dark blue) and $2\sigma$ (light blue). }
\label{fig:2d-lya}
\end{figure}

These two effects are calculated as follows.  The fractional change in the amplitude, $f_\Delta$, and the change in slope, $\delta n$, of the linear matter power spectrum for each $N$naturalness parameter point for $k_p$ which is defined by $z=z_p$.  These quantities are not very sensitive to the values of the $\Lambda$CDM parameters.  Recall the measurements made by eBOSS are $\Delta^2_{\rm lin}$ and $n_{\rm lin}$ which means the change induced by $N$naturalness at $z_p$ is estimated as
\begin{align}
\Delta^2_{\rm lin}(r,m_\phi) &= f_\Delta(r,m_\phi) \Delta^2_{\rm lin},\\
n_{\rm lin}(r,m_\phi) &= n_{\rm lin} + \delta n (r,m_\phi)\;.     
\end{align}
Figure~\ref{fig:2d-lya} shows the results.  The shaded blue regions are the $1\sigma$ and $2\sigma$ MCMC results from Planck while the solid (dashed) black line shows the $1\sigma$ ($2\sigma$) constraints from the Ly-$\alpha$ Gaussian likelihood (and does not include any other datasets).

For Dirac neutrinos, the constraints are similar to those obtained from the MCMC analysis that utilizes data both from the CMB and large-scale structure.  For Majorana neutrinos, on the other hand, the results are significantly stronger than those using CMB and large-scale structure data due to the substantial changes in the matter power spectrum at those wavenumbers.  The results demonstrate that Ly-$\alpha$ constraints have the potential to probe the $N$naturalness parameter space down to tuning at the sub-percent level.

We caution that Fig.~\ref{fig:2d-lya} is only an estimate of the impact of Ly-$\alpha$ data because the $\Lambda$CDM parameters are fixed to their best-fit values.  In an MCMC analysis, these parameters would be allowed to vary which we expect would relax the constraints.  Note also that the compressed eBOSS Ly-$\alpha$ dataset is reportedly in tension with the Planck dataset for the $\Lambda$CDM model because eBOSS prefers a strong suppression of the matter power spectrum around $k \approx 1h~{\rm Mpc}^{-1}$~\cite{Rogers:2023upm}. However, recent re-analyses of the eBOSS Lyman-$\alpha$ dataset with a narrower redshift range found it to be consistent with Planck data while fitting $\Lambda$CDM~\cite{Ivanov:2024jtl} (see also earlier study Ref.~\cite{Fernandez:2023grg}). Therefore, in this work,  we take a conservative approach and fit the $N$ naturalness-induced changes only to the error bar of the Ly-$\alpha$ likelihood.  Similarly, we anticipate that other smaller-scale measurements would also significantly probe the $N$naturalness parameter space for the Majorana case.
 
\section{Conclusions}
\label{sec:conclusion}

In this work, we have studied the cosmological signals of a tower of neutrino states in the $N$naturalness model, which addresses the long-standing Higgs hierarchy problem. Previous studies on the cosmological constraints of this model have primarily focused on its contribution to $\Delta N_{\rm eff}$ from additional $N$ species of photons.  In this study, we concentrate on the cosmological signatures of the additional sectors of neutrinos predicted by the model as a tower of warm DM states. Besides contributing to $\Delta N_{\rm eff}$, the presence of heavy neutrinos suppresses structure formation and introduces additional corrections to the CMB and matter power spectrum. Moreover, compared to scenarios with a single warm DM state, the multiple species of warm DM scenario results in a more gradual suppression of the matter power spectrum, transitioning from lower to higher $k$-modes.  The different neutrino states also introduce a redshift-dependence to $\Delta N_{\rm eff}$.

To make predictions of the cosmological signatures of $N$naturalness, we first derived the $N$-photon and $N$-neutrino mass and energy densities from the reheaton decay widths, showing that the $N$naturalness model is specified by only two parameters $(r, m_\phi)$ and the binary choice of Dirac or Majorana neutrinos.  After specifying the $N$-neutrino properties, we used \texttt{CLASS} to compute the CMB and matter power spectrum signals.  Studying constraints on higher $k$-mode perturbations requires the inclusion of a large number of neutrino states in \texttt{CLASS}. This presents a significant computational challenge, as solving the neutrino Boltzmann hierarchy is time-consuming due to its phase space evolution.  For each sector of neutrinos, we calculated the relativistic to non-relativistic transition in order to find the optimal truncation of the neutrino tower that does not sacrifice numerical accuracy.  In fact, this study motivates the future development of more efficient methods for describing multiple warm DM states, thereby speeding up the Boltzmann code for cosmological studies.

From the results, we show that the inclusion of $N$-neutrinos strengthens the bounds on the $N$naturalness parameter space. The $N$-neutrinos, especially in the Majorana case, produce significant suppression of the matter power spectrum at small scales. Therefore, incorporating small-scale datasets will further tighten the constraints on the model's parameter space. We derived estimated constraints from the compressed Lyman-$\alpha$ likelihood, which strongly probes the Majorana neutrino scenario.  These datasets also probe the case of Dirac neutrinos but to a lesser extent.

Although we have focused on the specific $N$naturalness model, this study provides valuable insights for other models with multiple warm DM states. Key features of the CMB and matter power spectrum, such as smoother suppression, can also appear in models where each warm DM state has different ``warmness" levels and transition times to non-relativistic particles. Inspired by Ref.~\cite{Dienes:2020bmn}, which investigates new ways to parametrize phase space distributions from multiple dark sector particles, it is beneficial to develop a systematic approach for identifying signals of multiple semi-relativistic particles, even with non-thermal energy distributions. With increasingly precise cosmological data, we can begin exploring the more complex structure of dark sector particles, potentially making links to profound mysteries at much higher energy scales, such as the Higgs hierarchy problem.

\section*{Acknowledgement}

We thank Kimberly Boddy, Marilena Loverde, and Linda Xu for the helpful discussions.  SG is supported by NSF grant number PHY-2112884.  SG and YT would
like to thank the Kavli Institute for Theoretical Physics (KITP) (supported by NSF grant PHY-2309135) for hospitality in the final stage of the work.  ML is supported by DOE grant number DE-SC0007914 and NSF grant number PHY-2112829. YT is supported by the National Science Foundation Grant Number PHY-2112540 and PHY-2412701. YT would like to thank the Tom
and Carolyn Marquez Chair Fund for its generous support. YT would also like to thank the
Aspen Center for Physics (supported by NSF grant PHY-2210452). This research was supported in part by the Notre Dame Center for Research Computing.

\appendix
\section{Parameter Tables and Triangle Plots}
\label{app:tables}

\begin{figure}
  \centering
  \includegraphics[width=1.0\textwidth]{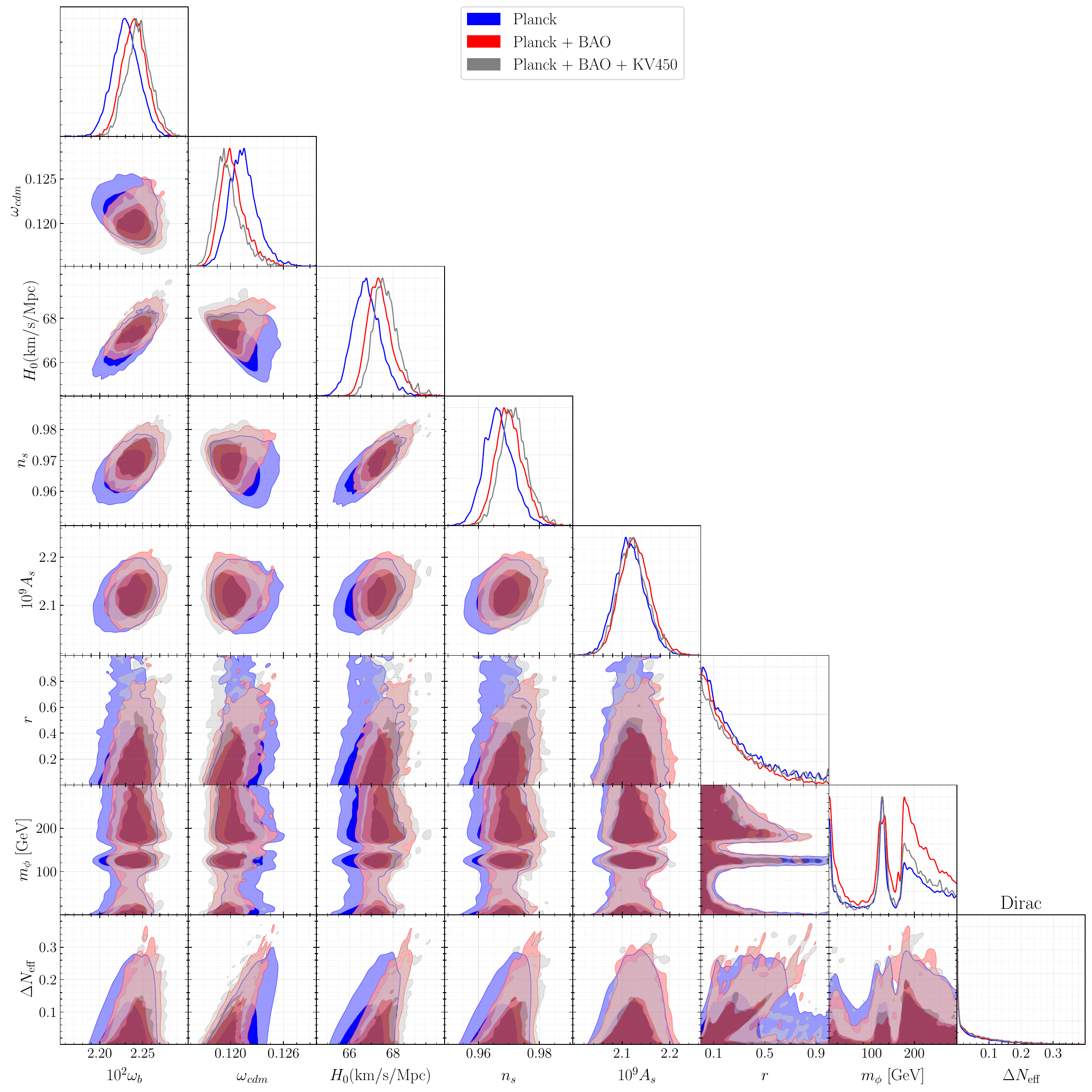}
  \caption{Triangle plot for $2D$ posterior distribution for several $\Lambda$CDM parameters and $N$naturalness parameters $r$ and $m_\phi$.}
  \label{fig:tri-dir}
\end{figure}

\begin{figure}
  \centering
  \includegraphics[width=1.0\textwidth]{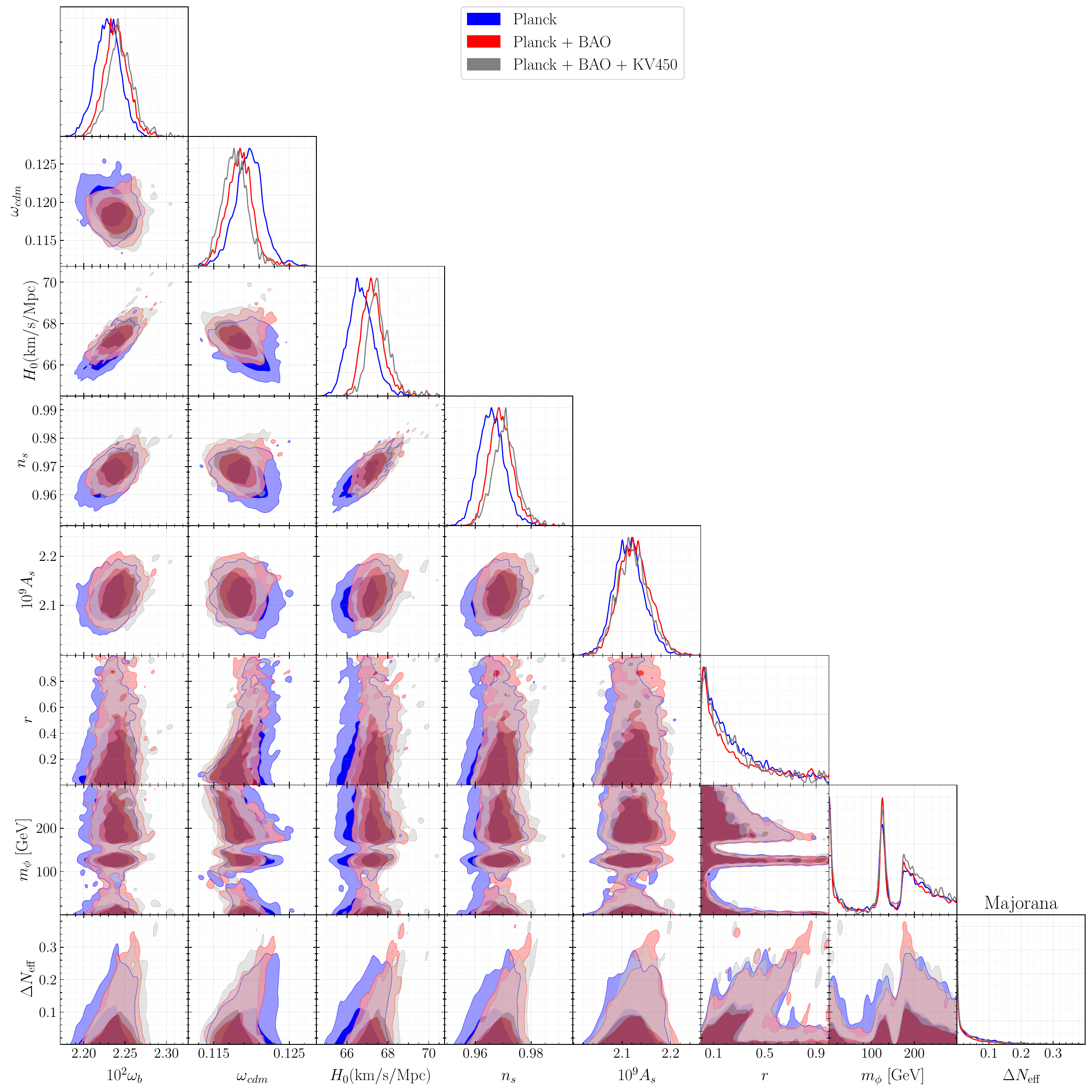}
  \caption{Triangle plot for $2D$ posterior distribution for several $\Lambda$CDM parameters and $N$naturalness parameters $r$ and $m_\phi$.}
  \label{fig:tri-maj}
\end{figure}

In this appendix, we provide the resulting parameter tables and the triangle plots from our MCMC analysis.  For Dirac neutrinos, the parameters are in Table~\ref{tab:param-table-dir} and the triangle plot is shown in Fig.~\ref{fig:tri-dir}.  For Majorana neutrinos, the parameters are in Table~\ref{tab:param-table-maj} and the triangle plot is shown in Fig.~\ref{fig:tri-maj}.

Figs.~\ref{fig:tri-dir} and~~\ref{fig:tri-maj} are useful both to understand some of the general trends of the $N$naturalness model and to highlight some of the differences between Dirac and Majorana neutrinos.  Starting with the $2D$ plots involving $r$, one can see there is a sizable region at small $r$ and a narrow region that extends to $r=1$.  From the $m_\phi$ vs. $r$ plot, we see that the distribution extending to $r=1$ comes from parameter points with $m_\phi \approx m_h$.  These correspond to the ``Higgs funnel'' where the mixing between the reheaton and the SM Higgs is maximal.  Consequently, the branching ratio into the SM always dominates over the branching ratio into any other sector irrespective of the value of $r$.

This behavior is also visible in the $r$ vs. $\Delta N_{\rm eff}$ plot where two branches are visible.  The branch at large $r$ with $\Delta N_{\rm eff}$ again corresponds to the Higgs funnel while the parameter points with larger $\Delta N_{\rm eff}$ are aggregated from the rest of the parameter space and indicate that generally increasing $r$ increases $\Delta N_{\rm eff}$ since the temperature of non-SM sectors increase with $r$.  These trends are present both for Dirac neutrinos and Majorana neutrinos.

One set of parameters that differentiates between Dirac and Majorana is the plot of $\Delta N_{\rm eff}$ vs. $\omega_{cdm}$.  For Dirac neutrinos, there is a small positive correlation between $\Delta N_{\rm eff}$ and $\omega_{cdm}$.  This arises because as $\Delta N_{\rm eff}$ increases to maintain consistency with data, the energy density of matter should also increase to prevent large changes to the time of matter-radiation equality~\cite{Planck:2018vyg}.  Dirac neutrinos have comparatively small masses and consequently contribute very little to the matter energy density which means the cold DM component needs to account for the increase in overall matter density.\footnote{Recall that the matter density is $\omega_m = \omega_{cdm} + \omega_b + \omega_\nu$.}  For Majorana neutrinos, on the other hand, the $N$-neutrino contribution to the matter energy density can be quite sizable and can often account for the required increase in matter energy density.  This is reflected in the $\Delta N_{\rm eff}$ vs. $\omega_{cdm}$ plot which is largely uncorrelated.

\renewcommand*{\arraystretch}{1.2}
\begin{table}
  \centering
  \begin{tabular}{|c|}
\hline
\\
\hline
Parameters \\
\hline
$10^2 \omega_{b}$\\
$\omega{}_{cdm }$\\
$H_0 ({\rm km/s/Mpc})$\\
$n_{s }$\\
$10^9 A_s$\\
$r$\\
$\tau{}_{reio }$\\
\hline
$\Delta N_{\rm eff}$\\
$\sigma_8$\\
\hline
\end{tabular}
  \begin{tabular}{|c|c|c|}
\hline
\multicolumn{3}{|c|}{Dirac}\\
\hline
Planck & Planck + BAO & Planck + BAO + KV450\\
\hline
$ 2.230\pm 0.016$ & $ 2.240\pm 0.014$ & $ 2.245\pm 0.015$\\
$ 0.1215^{+0.0012}_{-0.0018}$ & $ 0.12032^{+0.00095}_{-0.0017}$ & $ 0.11974^{+0.00092}_{-0.0018}$\\
$ 66.79^{+0.57}_{-0.68}$ & $ 67.40^{+0.44}_{-0.65}$ & $ 67.67^{+0.51}_{-0.60}$\\
$ 0.9666^{+0.0042}_{-0.0051}$ & $ 0.9701^{+0.0037}_{-0.0052}$ & $ 0.9716\pm 0.0045$\\
$ 2.117^{+0.028}_{-0.034}$ & $ 2.128^{+0.030}_{-0.033}$ & $ 2.122\pm 0.031$\\
$< 0.325$ & $< 0.271$ & $< 0.340$\\
$ 0.0550\pm 0.0076$ & $ 0.0585\pm 0.0075$ & $ 0.0580\pm 0.0072$\\
\hline
$< 0.0744$ & $< 0.0825$ & $< 0.0860$\\
$ 0.8018\pm 0.0062$ & $ 0.8003\pm 0.0062$ & $ 0.7976^{+0.0065}_{-0.0056}$\\
\hline
\end{tabular}
  \caption{Mean values and $68\%$ C.L. intervals (or $68\%$ C.L. upper limits from the $1D$ distribution) for several $\Lambda$CDM parameters and $N$naturalness parameters $r$ and $m_\phi$ for the Dirac neutrino case.  Note that the $m_\phi$ distribution differs greatly from a Gaussian distribution so the derived $1\sigma$ constraint on the distribution is not directly useful.}
\label{tab:param-table-dir}
\end{table}

\begin{table}
  \centering
  \begin{tabular}{|c|}
\hline
\\
\hline
Parameters \\
\hline
$10^2 \omega_{b}$\\
$\omega{}_{cdm }$\\
$H_0 ({\rm km/s/Mpc})$\\
$n_{s }$\\
$10^9 A_s$\\
$r$\\
$\tau{}_{reio }$\\
\hline
$\Delta N_{\rm eff}$\\
$\sigma_8$\\
\hline
\end{tabular}
  \begin{tabular}{|c|c|c|}
\hline
\multicolumn{3}{|c|}{Majorana}\\
\hline
Planck & Planck + BAO & Planck + BAO + KV450\\
\hline
$ 2.229\pm 0.015$ & $ 2.239^{+0.013}_{-0.015}$ & $ 2.244\pm 0.016$\\
$ 0.1196\pm 0.0019$ & $ 0.1184\pm 0.0017$ & $ 0.1178\pm 0.0017$\\
$ 66.68^{+0.56}_{-0.64}$ & $ 67.28^{+0.39}_{-0.60}$ & $ 67.57^{+0.46}_{-0.61}$\\
$ 0.9659^{+0.0040}_{-0.0050}$ & $ 0.9693^{+0.0035}_{-0.0048}$ & $ 0.9708^{+0.0039}_{-0.0046}$\\
$ 2.114^{+0.029}_{-0.032}$ & $ 2.126^{+0.030}_{-0.034}$ & $ 2.122^{+0.031}_{-0.040}$\\
$< 0.331$ & $< 0.346$ & $< 0.307$\\
$ 0.0544\pm 0.0075$ & $ 0.0585\pm 0.0075$ & $ 0.0581\pm 0.0074$\\
\hline
$< 0.0668$ & $< 0.0503$ & $< 0.0610$\\
$ 0.7956\pm 0.0088$ & $ 0.7955^{+0.0089}_{-0.0072}$ & $ 0.7919^{+0.0097}_{-0.0067}$\\
\hline
\end{tabular}
  \caption{Mean values and $68\%$ C.L. intervals (or $68\%$ C.L. upper limits from the $1D$ distribution) for several $\Lambda$CDM parameters and $N$naturalness parameters $r$ and $m_\phi$ for the Majorana neutrino case.  Note that the $m_\phi$ distribution differs greatly from a Gaussian distribution so the derived $1\sigma$ constraint on the distribution is not directly useful.}
\label{tab:param-table-maj}
\end{table}

\bibliographystyle{jhep}
\bibliography{refs}
\end{document}